# Resolution enhancement in quantitative phase microscopy: a review

VICENTE MICÓ [1, 6, *], JUANJUAN ZHENG [2, 6], JAVIER GARCIA [1], ZEEV ZALEVSKY [3], AND PENG GAO [2, 4, 5, *]

[1]*Departamento de Óptica y de Optometría y Ciencias de la Visión, Universidad de Valencia, C/Doctor Moliner 50, Burjassot 46100, Spain*
[2]*School of Physics and Optoelectronic Engineering, Xidian University, Xi'an, 710071, China*
[3]*Faculty of Engineering and the Nanotechnology Center, Bar-Ilan University, Ramat-Gan 52900, Israel*
[4]*Institute of Applied Physics, Karlsruhe Institute of Technology, 76131 Karlsruhe, Germany*
[5]*Institute of Nanotechnology, Karlsruhe Institute of Technology, 76344 Eggenstein-Leopoldshafen, Germany*
[6]*Contribute equally to this work*
*\* Correspondence: vicente.mico@uv.es (VM) and peng.gao@xidian.edu.cn (PG);*

Quantitative phase microscopy (QPM), a technique combining phase imaging and microscopy, enables visualization of the 3D topography in reflective samples as well as the inner structure or refractive index distribution of transparent and translucent samples. Similar to other imaging modalities, QPM is constrained by the conflict between numerical aperture (NA) and field of view (FOV): an imaging system with a low NA has to be employed to maintain a large FOV. This fact severely limits the resolution in QPM up to 0.82λ/NA, being λ the illumination wavelength. Consequently, finer structures of samples cannot be resolved by using modest NA objectives in QPM. Aimed to that, many approaches such as oblique illumination, structured illumination and speckle illumination (just to cite a few) have been proposed to improve the spatial resolution (or the space-bandwidth product) in phase microscopy by restricting other degrees of freedom (mostly time). This paper aims to provide an up-to-date review on the resolution enhancement approaches in QPM, discussing the pros and cons of each technique as well as the confusion on resolution definition claim on QPM and other coherent microscopy methods. Through this survey, we will review the most appealed and useful techniques for superresolution (SR) in coherent microscopy, working with and without lenses and with special attention to QPM. Note that, throughout this review, with the term "superresolution" we denote enhancing the resolution to surpass the limit imposed by diffraction and proportional to λ/NA, rather than the physics limit λ/(2$n_{med}$) being $n_{med}$ the refractive index value of the immersion medium.

*OUTLINE:*

# Resolution enhancement in quantitative phase microscopy: a review

**V. MICÓ, J. ZHENG, J. GARCÍA, Z. ZALEVSKY, AND P. GAO**

## 1. INTRODUCTION

Despite other high-resolution microscopy technologies, e.g., scanning electron microscopy (SEM), scanning tunnel microscopy (STM) and atomic force microscopy (AFM), optical microscopy acts as a mainstream inspection tool in the imaging community, especially, in biological studies. The main reason lies on optical microscopy is a minimally invasive technique and therefore compatible to live cell or tissue imaging. When being used with fluorescence specific labeling strategy, a certain kind of biomolecule on targeted structures can be observed. However, due to the diffraction-limit imposed by the wave nature of light, conventional optical microscopy has a spatial resolution around half of the used wavelength (typically ~200 nm) and thus, cannot resolve the subcellular structures finer than this limit. Since the 90s of the 20th Century, this limitation has been broken by the development of superresolution (SR) optical microscopy [1-3]. The key of SR microscopy is to switch single molecules "on" and "off" by light and therefore, a high-resolution image can be described by these single molecules. To cite a few, localization microscopy methods, including (fluorescence) photo-activated localization microscopy/stochastic optical reconstruction microscopy (PALM/STORM) [1, 4-6], stochastically excite single fluorophores that will appear apart from each other at a distance larger than the diffraction limit. Each fluorophore can then be separately localized (Fig. 1, left) with nanometer-scale resolution. In stimulated emission depletion (STED) [3, 9] [10-12] and reversible saturable/switchable optical linear (fluorescence) transitions (RESOLFT) microscopy [13, 14], a red-shifted, doughnut-shaped depletion light is superimposed with a Gaussian-shaped excitation light (Fig. 1, upper-right). The depletion light switches off the fluorophores except for its intensity minima, confining the fluorophores in the on state fluorescing in the reduced region. In addition, structured illumination (SI) microscopy (Fig. 1, upper-down) enhances the resolution by projecting fringes on the sample and recording the generated moiré patterns, and has clear merits as a tool for live cell imaging [8, 15].

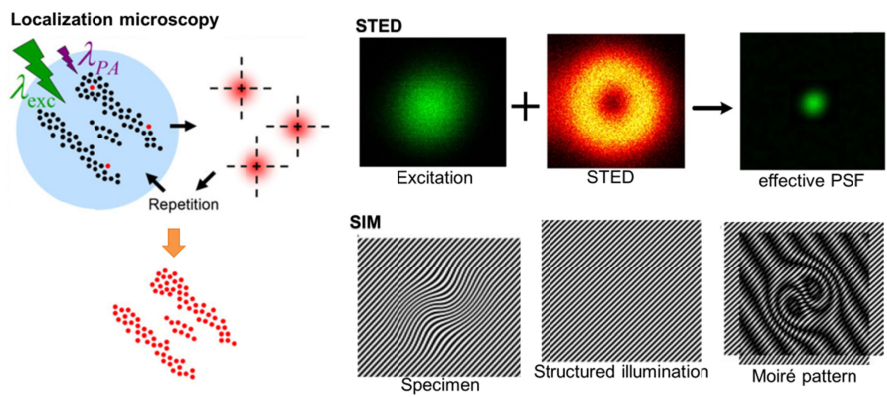

Figure 1. The state of art in SR fluorescence microscopy. The image on the left is taken from Ref. [7]. The image depicting SI microscopy taken from Ref. [8] - Copyright (2005) National Academy of Sciences, U.S.A.

There is no doubt that the usage of fluorescence markers has revolutionary advantages since it enables observation of specific targeted molecules. Nevertheless, it is also desirable to directly observe cells or tissues for their morphological structures in their natural condition, since fluorescence probes are harmful, or at least of disturbance, to biological systems. Of note, the determination of the 3D shape and refractive index of biological samples has recently experienced a growing interest in the biomedical community [16-18]. However, looking directly at biological samples is tremendously challenging since they are transparent or translucent under the visible light. Luckily, when a light propagates through a cell, its phase is distorted as a combination of both the thickness profile and the refractive index distribution of the cell. And this information can be exploited to visualize (qualitative imaging) and characterize (quantitative imaging) those transparent samples. For this purpose, phase contrast microscopy proposed by Zernike represented a major advance in intrinsic contrast imaging, as it revealed inner details of transparent structures without staining or tagging [19]. Holography concept was also conceived and implemented for phase imaging in the 60s and 70s [20-41]. However, both techniques were only qualitative. With the invention of the digital camera (CCD and CMOS) in the 90s, phase imaging approaches became commonly used techniques, and later becomes quantitative with the progress of digital analysis approaches. There came also other approaches for retrieving the phase information of a sample. To cite a few, wavefront sensing, employing a Shack-Hartmann sensor [42] or a pyramid sensor [43], is a simple and useful approach to investigate slightly distorted wavefronts. The basic idea is to spatially divide the tested wavefront into a series of sub-regions and focus them separately onto a CCD camera typically with a micro-lens array. Then, by analyzing the location of the spots on the CCD, the local slope of the wavefront over each region is computed. Alternatively, beam-propagation-based phase retrieval approaches can also retrieve the phase distribution by recording a series of diffraction patterns under different conditions or applying different perturbations [44-52]. For example, a beam diffracted by a phase sample influences the propagation of the beam and results in different intensity distributions at different planes. Vice versa, the phase can be recovered by analyzing the intensity distributions of a diffracted beam recorded at different distances [44], by using different illumination wavelengths [49-51], illuminating the object at different angles [53], applying a well-defined modulator [54], or varying the distance between the illumination source and the object [55].

In the application side, when a beam is reflected by the sample's surface or passes through a transparent sample, the phase of the beam contains information about, respectively, the 3D sample's profile or the refractive index distribution of the sample. This information can be exploited thus providing not only high-contrast images of transparent samples (qualitative improvement), but also quantitative evaluation on their 3D profiles or refractive index distributions [56]. The phase information can also be used for 3D imaging/display via holographic [20, 57] or integral imaging [58-60].

As an extension of phase imaging to microscopy, quantitative phase microscopy (QPM) [61-63] becomes a popular imaging choice when probing quantitatively biochemical and biophysical parameters of biological samples [64-68]. One appealing advantage is label-free, so that the studies with this technique do not require cells to undergo major preparation steps, and are targeted towards endogenous contrast. In that sense, QPM provides a versatile tool to observe or investigate the microscopic world quantitatively. Figure 2 summarizes the application of QPM, which includes industrial inspection [25, 70], special beam generation [71-74], air or liquid flow monitoring [75, 76], adaptive imaging [77]. In biology, QPM has been used to study refractive index, cell mass, mechanics, spectroscopy, shear stress, etc. [78-82]. QPM has also been used to investigate various cellular dynamics such as drug-induced changes [16] and laser-induced damage [83] in pancreas tumor cells. QPM has also been used to measure membrane fluctuation [63] and refractive index distributions [66, 68]. In contrast to conventional microscopy that observes samples in in-focus manner, QPM does not need it and is capable to digitally refocus the sample afterward by using numerical propagation tools. Hence QPM is suitable for 3D tracking of particles for sizing and velocimetry [84, 85]. Last but not least, QPM has also been used for suppressing the background noise in Coherent Anti-Stokes Raman Scattering (CARS) microscopy, considering resonant signal and non-resonant background has a phase shift π/2 [86, 87]. Similarly, nonlinear optical techniques, namely, harmonic holography have been developed for the coherent optical contrast between tracer particles and the background scatterings [88, 89].

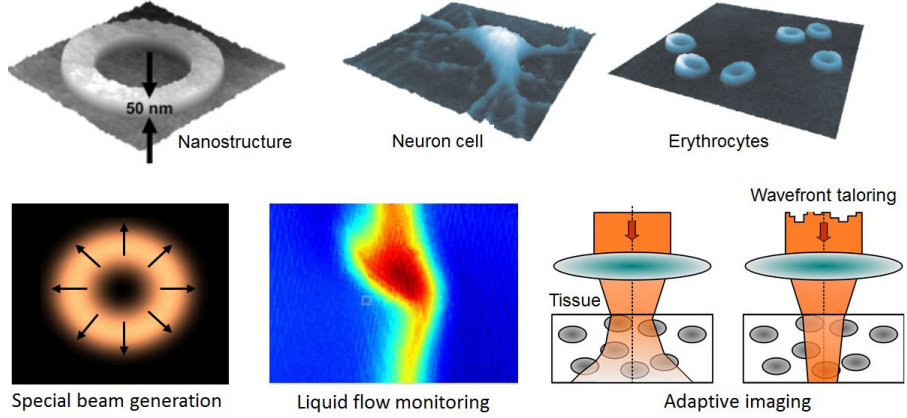

Figure 2. Applications of QPM. First image on the top taken from Ref. [25], and second/third images on the top taken from Ref. [69] (reprinted with permission from Lyncée Tec SA).

The resolving power is of great importance since it determines the minimal distance that two infinitely small spatial features can be positioned near each other and yet be seen as two separable items. Similar to other coherent imaging approaches, Illumination in QPM is provided by a coherent or partially-coherent light and the diffraction-limited resolution is defined as $\delta_{min}$ = 0.82λ/NA [90, 91] (*vide infra*, Appendix). Here, λ and *NA* indicate the illumination wavelength and the numerical aperture of the objective lens. In addition, the spatial resolution and the field of view (FOV) in QPM are closely related one to each other since the higher the former, the lower the latter (and vice versa). The ratio between these two parameters defines the SBP degree of freedom of the QPM imaging system [92]. The SBP concept generalizes the theorem of invariance of information capacity [93] because not only degrees of freedom are considered but also the shape of the SBP of the imaging system. Just as an example, a standard 20X microscope lens with a resolution limit of 0.8 μm and a circular FOV of 1.1 mm in diameter, provides a SBP of ~7 megapixels [53]. However, microscopists always expect to have higher SBP in the sense of increasing resolution while maintaining FOV or, in other words, to provide enhanced resolution under the same FOV. This can be accomplished from a theoretical point of view by matching the SBP of the sample with the SBP of the imaging system in order to allow effective transmission range of the sample's image through the microscope system [94, 95].

In the past decades, there were many efforts to improve spatial resolution without scarifying the FOV size. Resolution is simply one of several degrees of freedom that describe the imaging system. Though the total number of degrees of freedom is invariant, it is possible to sacrifice less desired ones, such as temporal, polarization or field-of-view constraints, to improve the final image resolution beyond the conventional diffraction limit. Some methods involve the use of shorter illumination wavelengths [96, 97], or by synthesizing a larger NA [98-101] using different illumination strategies such as oblique illumination (see Ref. [102] for an extensive reference list on synthetic aperture (SA) generation by tilted beam illumination), structured illumination [48, 103-108], grating projection [109-112], and illuminating with random patterns [113-117]. Depending on whether a reference wave is needed or not in QPM, these resolution enhancement approaches can be categorized into two groups: superresolved QPM via interferometric approaches and resolution enhancement by single-beam QPM.

**Table 1. List of acronyms in alphabetic order used in the paper.**

| Acronym | Full name description |
|---:|---|
| 1D | One dimensional |
| 2D | Two dimensional |
| 3D | Three dimensional |
| CCD | Charged coupled device |
| CMOS | Complementary metal-oxide-semiconductor |
| DC | Direct current |
| DH | Digital holography |
| DHM | Digital holographic microscopy |
| DIC | Digital interference phase contrast microscopy |
| DIHM | Digital holographic microscopy |
| FAL | First arriving light |
| FOV | Field of view |
| FPM | Fourier ptychographic microscopy |
| FT | Fourier transform |
| FWHM | Full width at half maximum |
| IRM | Interference holographic microscopy |
| LED | Light emitting diode |
| LHM | Lensless holographic microscopy |
| NA | Numerical aperture |
| ODT | Optical diffraction tomography |
| QPM | Quantitative phase microscopy |
| QPI | Quantitative phase imaging |
| PS | Phase shifting |
| RBC | Red blood cell |
| SA | Synthetic aperture |
| SBP | Space-bandwidth product |
| SEM | Scanning electron microscopy |
| SHG | Second harmonic generation |
| SIL | Solid immersion lens |
| SIIN | Solid-immersion imaging interferometric nanoscopy |
| SI | Structured illumination |
| SLIM | Spatial light interference microscopy |
| SLM | Spatial light modulator |
| SMIM | Spatially multiplexed interferometric microscopy |
| SNR | Signal to noise ratio |
| SPHM | Surface plasmon holographic microscopy |
| SPM | Surface plasmon microscopy |
| SR | Superresolution |
| SSM | Speckle scanning microscopy |
| SWI FPM | Surface wave illumination Fourier ptychographic microscopy |
| TLI | Turbid lens imaging |
| TIRHM | Total internal reflection holographic microscopy |
| USAF | United States Air Force |

This paper aims to complete and update previous manuscripts dealing with overviews about SR [102, 118] by providing an up-to-date review on the approaches to enhance the resolution in QPM. Moreover, it includes a survey about enhanced resolution in lensless QPM. This paper will also cover the pros and cons of each technique, as well as the confusion on resolution definition claim on QPM and other coherent microscopy techniques.

This article is organized as follows: Section 2 provides the theoretical foundations for understanding resolution enhancement in imaging systems; Section 3 reviews resolution enhancement approaches in reference-based interferometric QPM; Section 4 overviews SR approaches in reference-less QPM, including lensless holographic microscopy (LHM) field; Section 5 presents different ways for achieving SR imaging in QPM; and Section 6 summarizes this review. Finally, a description and some considerations about the criterion for evaluating the resolution limit in QPM is included in the Appendix and an extensive list of references concludes the paper. Also, Table 1 in Section 1 includes a list of all the acronyms used along the paper.

## 2. THEORETICAL BASIS FOR RESOLUTION ENHANCEMENT

### 2.1 Resolution limitation in QPM and SR concept

In conventional bright field microscopy, illumination is provided by an incoherent source (such as a thermal light) and the diffraction-limited resolution is defined as $\delta_{min} = 0.61\lambda/NA$. As previously stated, QPM has similar constraints in resolving power as that in conventional optical microscopy [$\delta_{min} = 0.82\lambda/NA$ [90, 91] (*vide infra*, Appendix)]. However, the resolving power of an imaging system is limited by several factors during the imaging process [118, 119]. Leaving aside the illumination wavelength, the first factor is the spectrum collection ability of the imaging system. The larger angle of the diffraction wave from the targeted sample structure can be collected, the higher resolution can be achieved. For the lensless QPM case, the spectrum collection is mainly limited by the aperture size of the CCD/CMOS camera for a given recording distance between a sample and the camera [Fig. 3(a)]. By contrast, for a lens-based imaging system, this factor depends mainly on NA designed of the objective used [Fig. 3(b)]. On both previous cases, the light diffracted from the sample beyond the angle defined by the effective aperture of the imaging system cannot pass through the system and therefore it is excluded in the image formation process [Fig. 3(c)]. The second factor which influences the resolving power of QPM lies in the performance of the CCD/CMOS camera. For the in-focus recording scheme in QPM, the pixel size is required to be two times finer pixel than one airy unit (defined by the diameter of one diffraction-limited spot after being imaged to the camera plane). However, for the out-of-focus recording scheme of QPM, a camera with larger space-bandwidth product (SBP) is required to record the spectrum of the object wave which spreads widely in the recording plane.

To evaluate the resolution enhancement in a quantitative manner, we assume the original numerical aperture (*NA*) of a QPM system is $NA_0$, and the increment of *NA* due to the enhancement of spectrum collection $NA_{incr}$, the effective *NA* can be formulated with:

$$NA_{eff} = NA_0 + NA_{incr} \tag{1}$$

To further consider the pixelated sampling of a CCD or CMOS camera, we denote the magnification of a QPM system with *M*, and the pixel size of the camera with $\delta$. Then, the eventual resolution can be written as:

$$\sigma_S = Max\left\{\kappa\frac{NA_{eff}}{\lambda}, \frac{2\delta}{M}\right\} \tag{2}$$

Here $Max\{\cdot\}$ denotes the operator to select the largest one from all the terms contained in the parenthesis. The coefficient $\kappa = 0.61$ for incoherence illumination and 0.82 for coherence illumination, but it is also related with other factors, such as, the signal to noise level of the image. Eq. (2) holds for imaging based QPM, while, for a Fresnel-based system finite-size of the light-sensitive area of the pixels attenuates higher spatial frequencies by convolving the reconstructed signal with a rectangular function of equal size to the light-sensitive area of the pixel [120].

For instance, for a 0.3 NA microscope lens, the cutoff frequency under on-axis coherent illumination due to diffraction is proportional to $0.3/\lambda$ [dash-gray circle in Fig. 3(d)], which defines a collection angle of 17.5 degrees in air (calculated from $NA_{obj} = 0.3$). Note that here we leave aside $k = 0.82$ for simplicity. Additional spatial-frequency information of the object's spectrum will be diffracted on-axis when the input object is illuminated with an off-axis beam [Fig. 3(c)]. By properly recovery and replacement of such circular frequency bands to its original position, a synthetic enlargement in the system's aperture will be achieved [Fig. 3(d)]. The spectrum collection is extended for both vertical and horizontal orthogonal directions where 2D oblique illumination strategy is considered. Theoretically, assuming the illumination and imaging $NA$ are, respectively, $NA_{illum}$ and $NA_{obj}$, the effective NA is $NA_{eff} = NA_{illum} + NA_{obj}$. And the cutoff frequency of the 0.3 NA system can be raised up near the theoretical limit ($1/\lambda$, under on-axis plane illumination) by illuminating the object with a set of oblique beams with $NA_{illum} = 0.7$. Moreover, it is possible to expand the synthetic cutoff frequency beyond the air-theoretical limit of $1/\lambda$ by considering additional oblique illuminations at higher angles $NA_{illum} > 0.7$. In this case, the synthetic cutoff frequency is expanded up to $(NA_{illum} + NA_{obj})/\lambda$ that can be higher than the air-theoretical limit depending on the parameters involved in the setup.

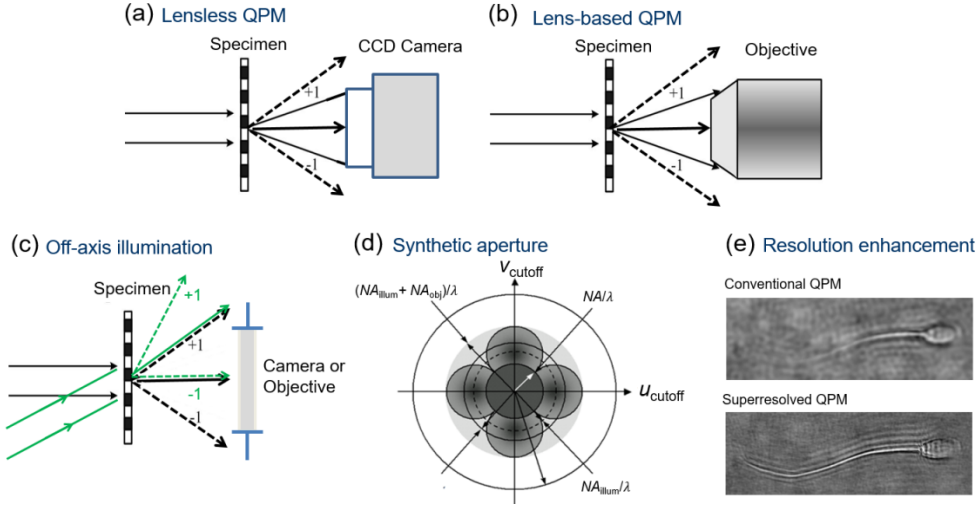

Figure 3. Space-frequency cut-off during QPM imaging. (a) frequency cut-off in lensless QPM and (b) in lens-based systems; (c) Scheme of off-axis illumination for resolution enhancement; (d) Space-frequency domain representation of the SA based on off-axis illuminations; (e) resolution enhancement of off-axis illumination based QPM. Up in (e), the phase image of a swine sperm cell obtained with conventional QPM; down in (e), the phase image obtained by QPM with off-axis illumination and SA generation. Images in (e) modified from Ref. [119].

We note that, here we take off-axis illumination as an example to explain the mechanism of resolution enhancement via aperture synthesis. Other resolution enhancement approaches, such as aperture synthesis in lens-less QPM, structured illumination and speckle illumination in lens-based QPM, share the same SA generation principle regarding resolution enhancement. In theory, the synthetic enlargement in the cut-off frequency of imaging systems in the air has a cut-off frequency limit of $2/\lambda$ (or $\lambda/2$ in resolution), corresponding to the highest angle at which diffracted light from the input object is available. This limitation can be achieved with an incident beam at grazing incidence ($\theta_{ilum} \cong 90°$) and diffraction in the backward direction ($\theta_{obj} \cong 90º$). In practice, the maximum *NA* of an air-immersion objective is limited to 0.95 and grazing incidence is not always possible and therefore, the theoretical cut-off frequency limit, $2/\lambda$, is reduced down to $1.9/\lambda$ due to practical constraints. A more detailed mathematical analysis reveal that this $2/\lambda$ cut-off limit is due to the definition of a transfer function associated with free propagation. We will deal in depth with this casuistic in Section 5.3 where evanescent field approaches for SR are included.

Recently, there are also several single image super-resolution (SR) techniques reported, which aims at recovering a high-resolution image from a single low-resolution image. For example, deconvolution [121, 122] enables to deblur a low-resolution image with known or partial information about the PSF. Moreover, machine and deep-learning methods recover a high-resolution image by exploiting internal similarities of the same image [123], or learn mapping functions from external low- and high-resolution exemplar pairs [124]. Model based measurement strategy pursues more information of the sample within a feedback loop combining the direct and the inverse principle [125]. These techniques can be formulated for generic image super-resolution, or can be designed to suit specific tasks, despite the single image super-resolution is inherently ill-posed since a multiplicity of solutions exist for any given low-resolution pixel [126]. In this review we mainly focus on the efforts which enhance the resolution from the hardware side, including but not limited to, the improvement of the illumination or detection scheme.

## 2.2. Degrees of freedom of an image

The previous analysis states that the $1/\lambda$ air-theoretical limit obtained in the conventional image with coherent illumination can be surpassed. People often assess the quality of an optical system solely in view of its spatial resolution. Nevertheless, the resolution is only one of the variables and parameters that determine the throughput of an optical system or transmission systems. The FOV, signal to noise ratio (SNR), the chromaticity transmission, and other parameters affect as well the system capability. Since the pioneering work of Shannon [127] dating back to 1948 the concept of quantity of information became a widespread measure of the capabilities of a system for dealing with data.

Before electronic and digital revolutions, the imaging systems produced analog images essentially static. The resolution of an optical system was known to be limited by the lens aperture size since the works of Abbe [128] on microscopy and Rayleigh [129] in telescope imaging. During the first half of the twentieth century, a new formalism was introduced for understanding optics in terms of Fourier decomposition, Fourier Optics [130]. In this paradigm, the resolution limit of a system could be understood as a fundamental limit given by the uncertainty principle between dual variables, namely, space and frequency. By the 50s this formalism was overlapped with the one of the emerging fields of electrical engineering, permitting a cross-breeding between both fields.

A landmark was the transposition of supergain antenna technology to optics. Toraldo di Francia first proposed that the spatial frequency cutoff of an optical system could be extended [131]. The underlying principle is to use a pupil that narrows the central diffraction spot at the expense of diverting light far from it, instead of trying to concentrate more the light in the central spot. The tradeoff for the enhanced resolution is a limited field and a decrease in the light efficiency. In practice, the reduction of energy that forms the image will result in a decrease of the SNR that makes this method impractical except for modest resolution gains [132]. These works show the link between spatial resolution, FOV, and SNR, showing how resolution can be enhanced at the expense of the other two parameters. This relation has been formalized and extended, again in close connection with electrical engineering developments by a number of authors [93, 131], [133-141]. Thus, Information theory serves to establish a relation between the resolution and the number of degrees of freedom of an optical system.

As early as 1955, Fellgett and Linfoot [134] derived the information capacity of a 2D optical system, showing it to be given by

$$incoherent\ case: N = 2L_xB_x 2L_yB_y \log\left(\frac{s+n}{n}\right)^{\frac{1}{2}}, \tag{3}$$

$$coherent\ case: N = 2L_xB_x 2L_yB_y \log\left(\frac{s+n}{n}\right). \tag{4}$$

Here $N$ is the number of degrees of freedom of the 2D optical system, $(B_x, B_y)$ are the spatial bandwidths, $(L_x, L_y)$ are the dimensions of the FOV in the $(x, y)$ directions, respectively, $s$ is the average power of the detected signal, and $n$ is an additive noise power. Note that the above expressions are just the product of the space-bandwidth product (SBP, giving the number of independent sampling points that represent the signal) and the number of bits that can be represented at each point, limited by SNR. Eqs. (3) and (4) implies that a coherent system has twice the information capacity of an incoherent system, while all the other parameters in Eqs. (3) and (4) remains equal. This may be understood by noting that in the coherent case each point is represented by a complex number given by two real values compared with the single one (intensity value) of the incoherent case.

Fellget and Linfoot's expressions for information capacity are also incomplete due to the absence of the temporal bandwidth. It is therefore desirable to develop an expression for the information capacity of an optical system that includes both the temporal and the spatial dimensions, and also noise and other degrees of freedom such as polarization states. In 1966, Lukosz proposed an invariance theorem to explain the concepts underlying all SR approaches [138], [140]. This theorem states that, for an optical system, it is not the spatial bandwidth but the number of degrees of freedom of the system ($N_F$) what is fixed. $N_F$ is, in fact, the number of samples needed to specify completely the system in the absence of noise and is given by

$$N_F = 2(1+L_xB_x)(1+L_yB_y)(1+TB_T), \tag{5}$$

where $T$ is the observation time, and $B_T$ is the temporal bandwidth of the optical system. Factor 2 in Eq. (5) is as a consequence of the two orthogonal independent polarization states.

This theorem not only quantifies the capabilities of the system, it can also be used to change the system design. Using this invariance theorem, that is, $N_F = constant$, Lukosz stated that any parameter in the system could be extended above the classical limit if any other factor represented in Eq. (5) is proportionally reduced, provided that some *a priori* information concerning the object will be known. For instance, if the spatial apertures of the imaging system are small, some of the spatial-frequency will be lost. If it is *a priori* known that the signal information is monochromatic, it is possible to convert part of the spatial information into wavelength information in such a way that the system's aperture results synthetically expanded. This theorem also supports the Françon proposal of a pinhole scanning method that can get arbitrary resolution, at the expense of long acquisition time [142] using a similar principle than near-field scanning optical microscope [143].

The Lukosz's number of degrees of freedom theorem meant an initial attempt to explain the underlying principles of SR in optical systems. However, this theorem is not complete because it does not consider noise in the optical system. Cox and Sheppard included the noise factor in the Lukosz's invariance theorem [143]. They considered the SNR as noise factor and also considered three spatial dimensions, two independent polarization states, and the temporal dimension

$$N = (1+2L_xB_x)(1+2L_yB_y)(1+2L_zB_z)(1+2TB_T)\log(1+SNR), \tag{6}$$

being $L_x$, $L_y$, $B_x$, $B_y$, $B_T$, $T$ defined in Eqs. (3-5), and $L_z$ is the depth of field, and $B_z$ is the spatial bandwidth in the $z$ direction. Factor 2 in each term of Eq. (6) considers the two independent polarization states of each independent dimension. It is worth noting the inhomogeneity of the degrees of freedom involved in the dimensional terms in the Eq. (6). We have terms connected with space and time while SNR deals with the amplitude of signals.

Frequently we are mainly concerned with spatial resolution and the time variable can be fully used for static samples. On the other hand, the logarithm in the SNR term often precludes its use, because a very large increase in signal (usually connected with energy) is needed to achieve a significant gain.

Then, the invariance theorem of information theory states that it is not the spatial bandwidth but the information capacity of an imaging system that is constant. Thus, provided that the input object belongs to a restricted class, it is in principle possible to extend the spatial bandwidth (or any other desired parameter) by encoding-decoding additional spatial-frequency information onto the independent/unused parameter(s) of the imaging system. Extreme examples of this application are the aforementioned pinhole-scanning imaging or the computational imaging [144, 145]. In this last technique, the image is encoded in a time sequence of scalar coefficients of base functions (using time as a recipient for spatial information), which, in addition, make use of the sparsity of the information of the image in the space-frequency domain.

As a final outcome, the resolution of an optical system is no longer understood as a separate variable, but it is related with all the variables that can carry information. Thus, a constant volume of information is defined that can be squeezed to increase the range of a given parameter at the expense of compressing any other. This is the fundamental principle underlying all SR schemes, which are an ingenious way to reshape the original information volume of an optical system so that it will be elongated in the direction of the resolution.

### 2.3. Space bandwidth product adaptation

In addition to provide a mathematical basis for enhancement of resolution limits, the above-presented approach also includes a powerful method for visualizing the information content of an image or the information capacity of a system. Centering on spatial information content, the information content of an image is proportional to the area it covers in a space-frequency representation, such as Wigner distribution [146], determined by the bandwidth $\Delta v$ and the object extent $\Delta x$ [Fig. 4(a)], the product of both defining the space-bandwidth product (SBP). The SBP of a signal is defined by the spatial location and by the range of spatial frequencies within which the signal is nonzero. Then, the SBP may be either a pure number (degrees of freedom in the case of the imaging system) or a specific area in the spatial-frequency domain (Wigner domain for the optical signal case). The SBP concept generalizes the theorem of invariance of information capacity because not only degrees of freedom are considered but also the shape of the SBP of the imaging system.

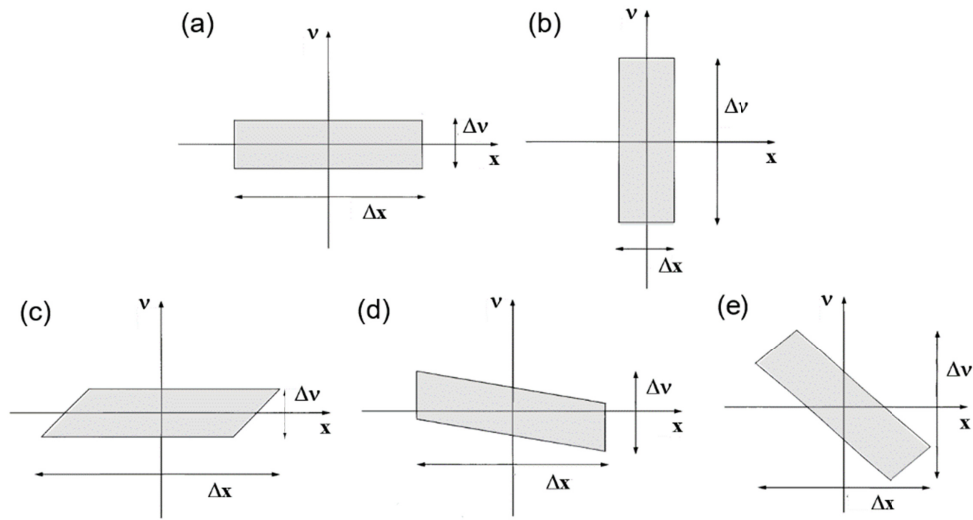

Figure 4. Space-frequency representation in Wigner space; (a) SBP distribution of a sample image (b) SBP of the FT of the signal (c) effect of free space propagation (d) Effect of the transmission through a lens. (e) effect of a fractional FT. Figures adapted from Ref. [92].

The common optical operations imply an affine transformation in this representation. The Fourier transformation (FT) operation rotates the distribution by 90 degrees, mapping the former frequencies in space coordinates [Fig. 4(b)]. Free space propagation and the action of a lens produce a shearing along the $x$ axis or the $v$ axis, respectively [Fig. 4(c)-(d)]. We can also define an arbitrary rotation, which can be obtained as a combination of the previous operations and is defined as a fractional FT [147]. Note that these operations preserve the area of the SBP. Optical magnification (compression in spatial axis along with expansion in the frequency axis) or the action of a prism (shift in the frequency axis) will also maintain invariant the SBP area. Other operations, like the multiplication by another function or the effect of aberrations can change the support of the SBP.

Mendlovic et al. introduced the SBP adaptation concept as a tool for matching signals with specific systems and with application to SR [92], [148], [149], [150]. The SBP might describe either an optical system or an optical signal. From the optical system point of view, it may serve to perform imaging, and The SBP can be applied to SR by considering that the SBP of the optical signal is well adjusted with the SBP of the imaging system. The key point is to match the SBP of the optical signal with the SBP of the imaging system to allow effective transmission range of the optical signal through the imaging system.

It is interesting to point that with the generalization of the use of solid state sensors and digital computing in optics, the space bandwidth representation and its connection to sampling theory takes a direct meaning, as most optically recorded images are inherently sampled and the SNR of each sample limited by the sensor quantization (aside other factors, of course). A deep understanding of these concepts is the key for designing efficient algorithms for the calculation of optical transformations, making use of the transformation of the SBP of the signal along an optical system [151, 152].

Understanding the system requirement and limitations for digital holography (DH) also profits from the visualization of SBP of the signal and the corresponding hologram. Figure 5 shows the original signal SBP along with the space-frequency schematic representation of diverse types of holograms. We can observe that the original signal [Fig. 5(a)] produce different spatial and frequency extension when recorded as a hologram. We should consider the shearing of the SBP of the image due to the introduction of the reference beam in on-axis [Fig. 5(b)] and off-axis arrangements [Fig. 5(c)]. Here, the DC term of the hologram and the replicas introduced by the interference with the reference and the subsequent intensity recording provide different SBP representations where it can be clearly seen as the inline holograms have the lowest requirements in terms of sensor space-bandwidth capability.

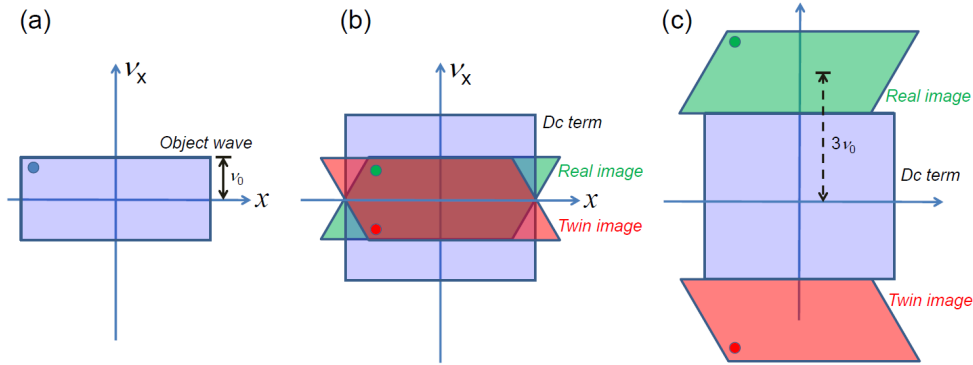

Figure 5. SBP representation for different Fresnel-type holograms in Wigner space: (a) original SBP of the object, (b) SBP for in-line geometry, (c) SBP for off-axis geometry without suppression of DC term, $v_0$ indicates the maximal frequency of the object wave. Figure adapted from Ref. [153].

Note that we are considering in all cases far-field imaging, meaning that evanescent waves propagating from the object are attenuated to a level that makes their recovery unfeasible for any conceivable SNR. Direct methods, such as near-field microscopy [143], imaging of evanescent waves [154] or negative index material imply contact or almost contact (distances similar or smaller than the imaging wavelength) to couple the evanescent wave into the imaging system. The formalisms derived for propagating waves (far-field) can usually be used without significant changes also to SR over the absolute limit given by the wavelength.

## 3. RESOLUTION ENHANCEMENT APPROACHES IN REFERENCE-BASED QPM

### 3.1. DHM with oblique illumination

Digital holographic microscopy (DHM), a method that combines digital holography and microscopy, is one of QPM variations. In a regular lens-based DHM where a microscope lens is used to provide a magnified image of a specimen which is illuminated by a non-tilted plane wave, only the spatial-frequencies diffracted by the sample and up to a maximum value derived from the NA (~NA/λ) of the objective will be transmitted through the limited aperture of the microscope lens (Fig. 6, upper part). The high spatial frequency content is diffracted at higher angles falling outside the limited lens aperture. However, when a plane but oblique wave is used as illumination (Fig. 6, lower part), part of this higher spatial-frequency range is diffracted on-axis and, thus, will pass through the limited lens aperture. This downshift of the high spatial frequency range is recovered in a similar way than the centered spectrum aperture by DHM does. After multiplexing different incidence directions, a complementary set of apertures containing different spatial-frequency content is available and a SA is generated as the coherent addition of such aperture set. Figure 6 depicts the addition of 4 external apertures for SA generation.

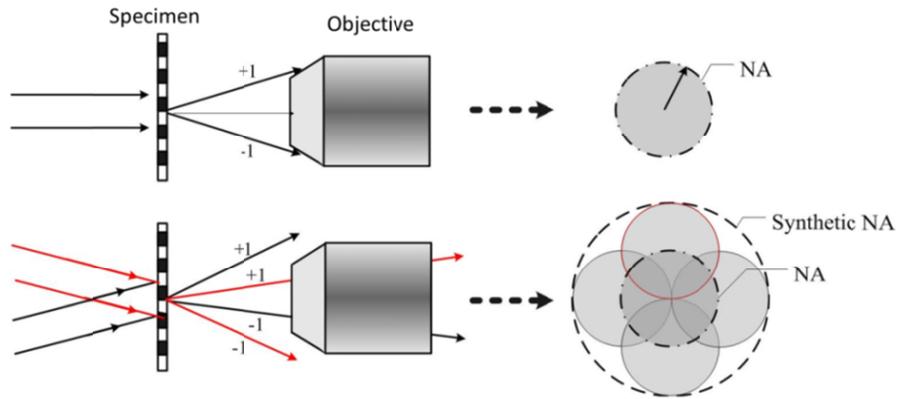

Figure 6. Schematics of oblique illumination and SA generation.

The reconstruction process is done digitally and the downshifted frequencies are back assembled to their original positions in the object's spectrum, thus synthesizing a wider spectrum than the one defined by the regular limited aperture. The final superresolved image is achieved by inverse FT of the generated SA. It is worthy to mention that the SA synthesis is a coherent process where different phases need to be equalized and properly adjusted. These aberrations come from different recording conditions (slight mismatches in the experimental layout) of the different holograms recorded with different oblique beams. Thus, the set of recovered elementary apertures must be carefully managed to provide a high-quality reconstructed image. For this purpose, the phase difference of each object wave (for example, the ith object wave) under the oblique illumination with respect to the on-axis illuminated object wave is calculated, fitted with low-order 2D polynomial function, yielding $\varphi_{abber}(x, y)$. Then, the ith object wave is multiplied by $\exp(-i\varphi_{abber})$ to compensated for the aberration. Figure 7 shows the significance of the proper coherent matching of the recovered elementary apertures. The highest image quality can only be obtained when the phase offset between different object waves are removed [Fig. 7(c)].

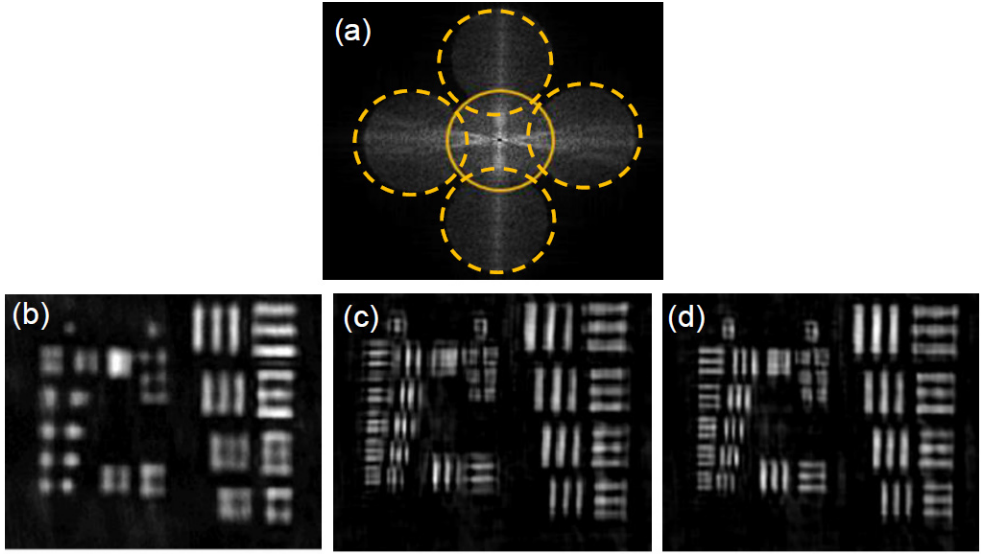

Figure 7. The significance of phase matching of the recovered elementary apertures during aperture synthesis. (a) Spectrum of conventional QPM image versus that of SA QPM image containing 4 additional apertures (dash-yellow circle) into the horizontal and vertical directions at the Fourier domain (the solid-yellow circle indicates the cut-off frequency limited by the numerical aperture of the system), (b) conventional low-resolution image, and (c-d) the superresolved images with and without phase matching on the different elementary apertures.

This phase matching procedure can be guided by essentially two different types of criteria: i) mathematical optimization of a given metric from the overlapped regions coming from each elementary aperture, and ii) reconstructed image quality based criteria. The former yields in a quantitative criterion where the position of each coherent aperture in the Fourier domain is typically commanded by computing cross-correlation operation between the overlapping areas in adjacent recordings [155-158] or other metric optimizations such as the sharpness metric [159, 160], the spectrum normalization [161] or a mean-square-error metric [162]. In these cases, the reconstruction becomes more accurate as the overlapping area increases but a higher number of recordings are needed to get a reasonable resolution gain factor. For this reason, there are other criteria based on the final quality of the reconstructed image. Some procedures are based on maximization of both the reconstructed image quality using image variance criterion and the spectral energy distribution ratio between each pair of adjacent holograms [163] or by visual evaluation of the reconstructed image quality every time that a new band-pass image is added to the others [164, 165].

Using oblique illumination strategy it is possible to generate a synthetic image containing sub-resolution information. Here sub-resolution means that it contains details incoming from the spatial frequencies that are outside the conventional aperture under plane wave illumination. This is the reason why this type of techniques for resolution enhancement are also named SR methods because it is related with the capability to overcome the resolution limit imposed by diffraction without changing the NA value of the lenses. So, given that the air immersed resolution limit in DHM under plane wave illumination is $0.82\lambda/NA$, it is improved to $0.82\lambda/[NA+NA_{ilum}]$ by using oblique beam illumination. Although it is possible to reach the theoretical limits, this type of approaches are commonly validated for regular imaging systems with modest spatial resolution (NA up to 0.5 range, approx.).

Since the first implementations of SR imaging in QPM by oblique beam illumination [166-169], many approaches to synthesize a larger aperture have been extensively reported in the literature [67, 157, 162, 165, 170-186]. Schwarz et al. used off-axis illumination to downshift the high-frequency components of the object's spectrum, and a reference beam is reinserted just at the proper angle to upshift the image content back to its proper spatial frequency region [101, 162]. This process was performed twice for orthogonal illumination directions and sequentially in time, thus allowing the transmission of different frequency bands of the 2D object's spectrum. The authors obtained a resolution gain in a factor of 3 for both orthogonal directions. Subsequent works demonstrated resolution improvement until the maximum limit for air-immersed optical imaging systems [162] as well as considering evanescent waves [170]. Also with optical rearrangement of the additionally transmitted spectral ranges, Mico et al. reported on the use of a vertical-cavity surface-emitting laser (VCSEL) array for SR [185]. In this case, the resolution improvement was 1D but it was obtained with a single camera snap-shot with a resolution gain factor of 5. Extension to 2D imaging was achieved using different interferometric configurations [165, 180, 183, 187]. Notably, synthetic NA above the maximum limit for air immersed imaging systems (NA = 1) was also reported [119]. Additional implementations considering wavelength multiplexing [174-176], and adaptation of oblique beam illumination techniques have been recently implemented into a regular upright Olympus microscope [179]. The use of scanning elements has also been reported as a technological improvement for providing oblique beam illumination [162, 173, 177, 188], and sample rotation instead of tilt the illumination direction was also validated [111, 170, 172].

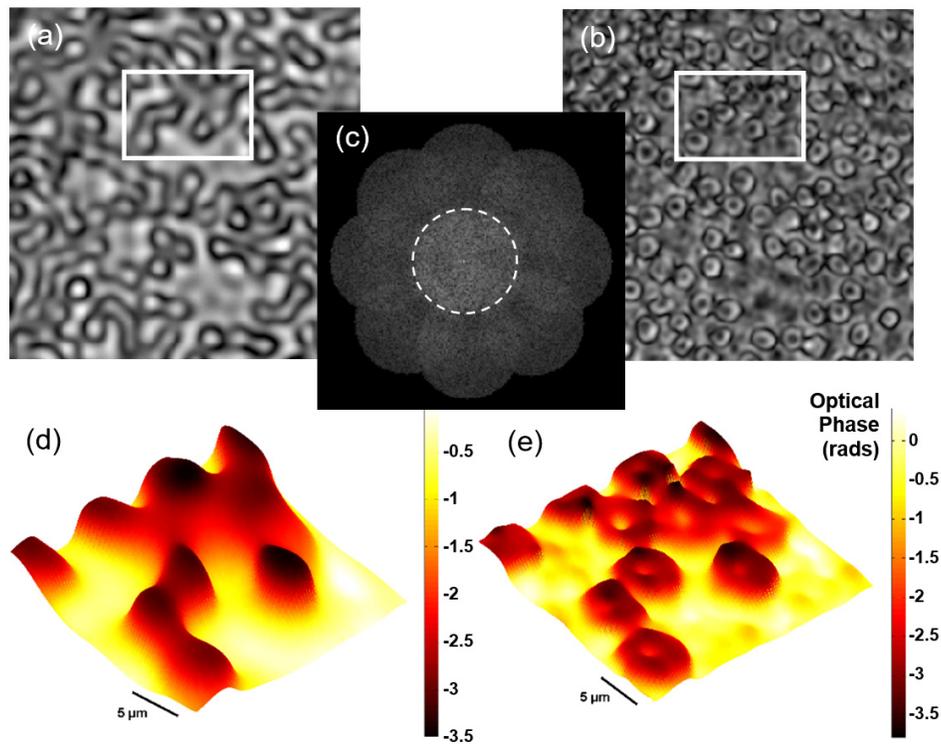

Figure 8. Resolution-enhanced QPM imaging using oblique beam illumination for a human RBCs bio-sample. (a) the conventional (low resolution) image, (b) the superresolved image,

(c) the SA in comparison with the conventional one, and (d)-(e) 3D plots of the ROIs in (a)-(c), respectively, showing quantitative phase information of the RBCs. Images taken from Ref. [171].

Oblique beam illumination for resolution enhancement has also been implemented by using a spatial light modulator (SLM) [189-191] as well as by multi-fiber based illumination [192-196]. Finally, it has also been applied to differential interference contrast (DIC) microscopy [197] and Zernike phase contrast microscopy [198]. Moreover, axial rather than transversal resolution improvement has also been validated by the SA generation [199] and an application of the technique to edge processing was also reported [200].

Just as an example to demonstrate resolution enhancement by the oblique beam illumination technique, Fig. 8 shows the experimental results provided by the method reported in Ref. [171]. 0.27 Synthetic NA was obtained for a microscope objective with 0.1 NA while 2D full space coverage of the sample's spectrum via SA [shown in Fig. 8(c)] enables RBCs visualization [Fig. 8(b)] with high resolution in comparison with the low-resolution images provided by the objective under conventional on-axis illumination [Fig. 8(a)]. Moreover, superresolved quantitative phase imaging (QPI) is available since complex amplitude distribution is retrieved. This fact is demonstrated through Figs. 8(d)-(e) where a 3D view of the optical phase is represented.

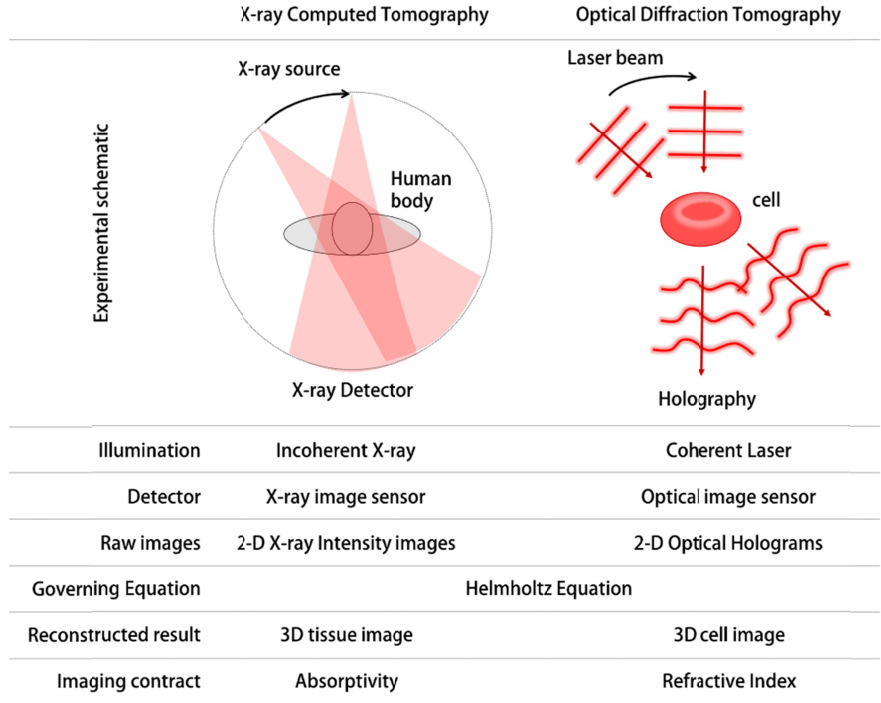

Figure 9. Comparison between X-ray coherence tomography and optical diffraction tomography. Image taken from Ref. [202].

It is interesting to mention that the oblique illumination can also be used to obtain 3D refractive index (RI) tomographic images of transparent or translucent samples via optical diffraction tomography (ODT). ODT was firstly proposed by E. Wolf [201] and has recently demonstrated for 3D imaging of live cells. In ODT (Fig. 9), multiple 2D images will be taken from the sample by using different oblique illuminations. Then, a 3D refractive index distribution is calculated by using the filtered back-projection method [203, 204], analogous to X-ray coherence tomography, in which 3D absorption distribution is calculated. The form in Fig. 9 summarizes the differences between the two techniques. Furthermore, the diffraction of light induced by the sample can be used to take into account with the reported algorithms [201, 205, 206]. In the earlier time, ODT was often performed with interferometric approach [81, 203] by introducing an independent reference wave, and later ODT was also performed with single-beam phase retrieval approach [207]. Theoretically, the spatial resolution of ODT is defined as the maximum spatial frequency in reconstructed Fourier spectra [202]. For the ODT realized by rotating the illumination beam, resolution enhancement on both lateral and axial direction can be achieved due to the "SA" of different illumination beams. Recently, several studies applied ODT in the investigation of the physiology of biological cells, including RBCs, white blood cells, and various eukaryotes. It is worthy to note that, in these applications, refractive indices can be quantitatively and precisely measured at the individual cell level [208], whereas conventional complete blood count machines measure averaged values from many cells.

### 3.2. DHM with structured illumination

As mentioned before, SI microscopy allows for observation of fluorescent samples at resolutions below the diffraction limit by projecting fringes on the sample and recording the generated moiré patterns, and has clear merits as a tool for live cell imaging [8, 15]. But SI has also been applied for non-fluorescent, coherently scattering samples [103, 112, 209, 210] and, over the last years, SI has also been combined with DHM to image transparent samples with resolution enhancement [104-108, 211-217]. Figure 10(a) shows a schematic setup used for the DHM with SI [211]. In the implementation, four binary phase gratings rotated by m×45° [Fig. 10(b)] and generated by a SLM are projected onto the sample, so the sample is illuminated sequentially by 4 sinusoidal fringe patterns. Here $m$=1, 2, 3, 4 indicating different orientations of the fringe illumination. After passing a telescope system comprised by the microscope objective $MO$ and the lens $L$, the object wave interferes with a tilted reference wave $R$ and the generated holograms are recorded by a CCD camera. $\Phi_{mn}$ denotes the SI wave generated by the grating having $m$-th orientation and $n$-th phase shifting (PS), and with $\Psi_{mn}$ the resultant wave transmitted through the object when becomes illuminated by $\Phi_{mn}$. Thus, $\Psi_{mn}$ interferes with the tilted reference wave $R$ on the detector plane generating an intensity hologram in the form of $I_{mn} = |R + \Psi_{mn}|^2$. From this intensity, the wave $\Psi_{mn}$ can be reconstructed by using standard methods as in off-axis DHM. Generally, $\Psi_{mn}$ can be decomposed into three waves $A_{m,-1}$, $A_{m,0}$ and $A_{m,1}$ along the -1st, 0th and +1st diffraction orders of the illumination wave. When we assume that the phase increment for each shifting of the grating is $\alpha$, then $\Psi_{mn}$ can be written as:

$$\Psi_{mn} = \gamma_{-1}\exp(-in\alpha)A_{m,-1} + \gamma_0 A_{m,0} + \gamma_1 \exp(in\alpha)A_{m,1} \tag{7}$$

where $\gamma_{-1}$, $\gamma_0$ and $\gamma_1$ denote the magnitudes of the diffraction orders. From Eq. (7) we can calculate the reconstructed object waves $A_{m,-1}$, $A_{m,0}$ and $A_{m,1}$ coming from the different diffraction orders and combined in the Fourier plane to yield the synthetic spectrum as:

$$\begin{bmatrix} A_{m,-1} \\ A_{m,0} \\ A_{m,1} \end{bmatrix} = \begin{bmatrix} \gamma_{-1}\exp(-i\alpha) & \gamma_0 & \gamma_1\exp(i\alpha) \\ \gamma_{-1}\exp(-i2\alpha) & \gamma_0 & \gamma_1\exp(i2\alpha) \\ \gamma_{-1}\exp(-i3\alpha) & \gamma_0 & \gamma_1\exp(i3\alpha) \end{bmatrix}^{-1} \cdot \begin{bmatrix} \Psi_{m1} \\ \Psi_{m2} \\ \Psi_{m3} \end{bmatrix} \tag{8}$$

Finally, a focused image with enhanced resolution is retrieved by inverse FT of the synthetic spectrum. As in previous oblique beam illumination approaches, the enhanced resolution limit in DHM with SI is determined by the illumination angle $\theta_{\text{illum}}$ of the +1 and -1 diffraction orders. Assuming the angular aperture of the imaging system (limited by the objective $MO$) is $NA_{MO}$, the synthetic $NA$ of the SI DHM is: $NA = NA_{MO} + \sin\theta_{illum}$ [Fig. 10(c)]. And this synthetic NA means an enhancement in spatial resolution, compared with the on-axis plane wave illumination where $\theta_{illum} = 0$. Specifically, in the experiment [211] the $NA_{MO} = 0.25$ of the objective lens limits the resolution of the set-up to $\delta_{plan} = 1.55\mu m$ for the on-axis plane wave illumination. Using the SI (with $\theta_{illum} = 10$ degrees), the resolution was improved to $\delta_{str} = 0.9$ μm. Particle-cluster on a glass plate (amplitude and phase object) have been used to validate the resolution enhancement of the described method (Fig. 11). The phase images obtained by an on-axis plane wave and SI are shown in Fig. 11(a)-(b), respectively. The phase image in Fig. 11(b) has a better resolution compared with that in Fig. 11(a) and two particles separated by 1.4 μm become distinguishable as shown in Fig. 11(c). The full width at half maximum (FWHM) of a single particle is 1 μm, which is slightly larger than the theoretical resolution value. We note that the SI allows for resolution enhancement in the axial direction, in aside of the lateral directions [218]. This is due to the fact that, the on-axis illumination in conventional DHM has infinitesimal axial frequency support, while thoroughly coverage of 3D frequency-space can be obtained by using structured illumination with different spatial frequencies. Therefore, similar to angle-scanning optical diffraction tomography [202], structured illumination based DHM can enhance the axial resolution. In the meantime, the out of focus component, which is barely modulated by the structured illumination, can be removed by the phase shifting of the illumination fringe, contributing to further axial resolution enhancement.

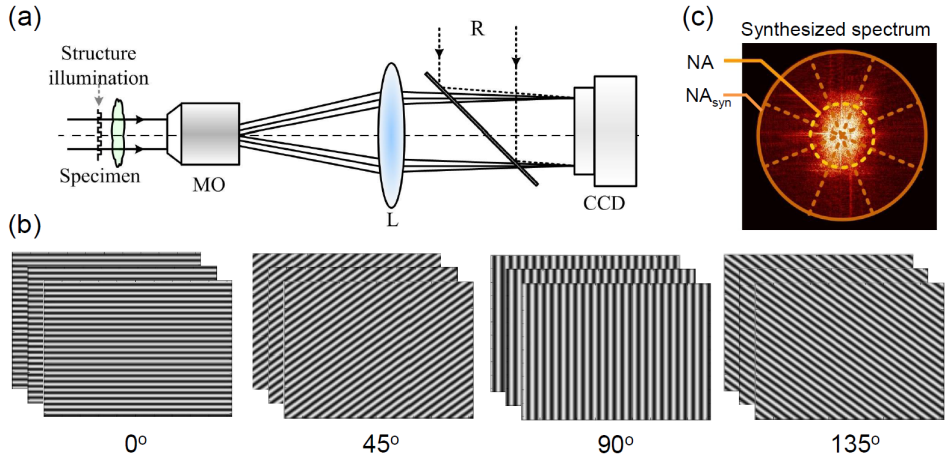

Figure 10. DHM with SI. (a) Schematic setup; (b) four groups of SIs with different directions; (c) the synthesized spectrum. Images taken from Ref. [211].

Despite there are some similarities between SI-DHM and SI microscopy reconstruction, there are also some differences between the two. In SI microscopy, the measured intensity distribution at the image plane is a linear transform of the fluorescent emission distribution at the sample, and thus SR reconstruction requires only the measurement of the intensity. In SI-QPM, the sample is illuminated with a coherent field, and the resolution enhancement of both the amplitude and phase images requires the measurement of the complex amplitude of the field. Moreover, the SI in DHM is completely analogous to the case of simultaneous illumination with two oblique plane waves at two different angles and provides the same effect achieved by illuminating first at one angle and then at the other one. The reconstruction process is then essentially a separation of these multiplexed components, which is analogous to properly oriented illumination beams [217]. By contrast, in conventional SI microscopy, the achieved resolution gain cannot be attributed to any single beam, regardless of orientation.

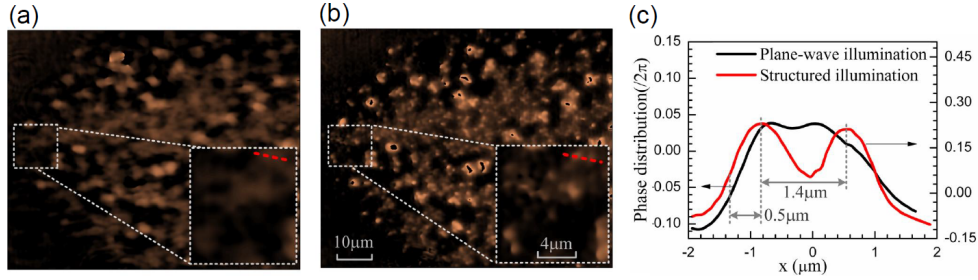

Figure 11. Experimental results for resolution enhancement in DHM with SI. (a) and (b) reconstructed phase images by using plane wave illumination and SI; (d) the phase distributions along the two lines drawn in (a) and (b). Images taken from Ref. [211].

Recently, SI with unknown sinusoidal SI has been exploited for DHM, for which an iterative reconstruction is needed to obtain resolution-enhanced images [106, 107, 216]. SI has also applied for multimodal sub-diffraction imaging of both coherent quantitative-phase and fluorescence [218, 219]. SI has also been applied in DIC microscopy [220] and optical coherence tomography [221]. Notably, the later one, termed as structured interference optical coherence tomography (SIOCT), is a three-dimensional imaging modality with micrometer-level resolution and millimeter-level penetration depth. In SIOCT, the illumination is implemented by a confocal scheme that allows a focused Gaussian beam to scan laterally. The reflection from the sample interferers with the reference wave, and the interference field is dispersed and recorded by a spectrometer for axial determination. In SIOCT, a sinusoidal pattern is created on the interferometric beam by temporally modulating the reference intensity. A sophisticated reconstruction provides a two-fold enhancement on the lateral resolution. We note that SI-DHM allows phase imaging of non-fluorescent samples at a resolution level up to twice the diffraction limited resolution $\sim\lambda/NA$ (limited by a specific microscope objective), but it does not contain more information beyond the physics diffraction limit, namely, $\sim\lambda/2$ [222].

### 3.3. DHM with speckle illumination

The speckle phenomenon has long been familiar especially after the introduction of the laser in 1960 [223]. Generally, speckle fields can be classified into fully developed speckles and partially developed speckles. The prior one is described as the result of random phasor sums, and the second one is described as a constant phasor plus a random phasor sum [224, 225]. Speckle techniques were exploited for studying displacement and deformation as well as vibration and stress analysis via speckle pattern photography, digital speckle correlation method, and electronic speckle pattern interferometry, speckle shearography [226]. In incoherent microscopy, the speckle patterns generated by a rotating diffuser provides sectioning capability similar to that of confocal microscopy, and in the meantime, it contributes to improving spatial resolution by utilizing full condenser aperture [227]. Despite unwanted diffraction can be reduced by averaging different images under different speckle illumination, the phase information is lost during the time-averaging of the speckles. In order to retrieve phase information, each speckle field (including the information of both the targeted sample and the illumination pattern) should be recorded by interfering with a reference wave [113, 228-230].

Recently, speckle fields were also used to enhance the resolution of phase imaging, and to reduce the coherent noise in DHM [113, 116, 177, 230]. Figure 12 depicts the experimental setup included in Ref. [230] and involving sequential speckle illumination. In the object wave path, a holographic diffuser is used to generate speckle field illumination. A galvanometric mirror (GM) before the holographic diffuser controls the incident angle of the laser beam onto the diffuser and generates angle-dependent speckle fields. Two sets of angle-dependent speckle images were recorded, one with the sample and the other without it. And two sets of object waves with and without the sample are reconstructed from the phase-shifted DHM holograms, which are obtained by utilizing the acoustic optic modulators ($AOM_1$ and $AOM_2$). Then, a set of angle-dependent electric-field images were obtained after dividing one set with the sample by the other without it and the sample-induced complex field image is retrieved. However, along with each speckle field, there are singular points (vortex) in the resulting amplitude and phase maps. To solve this problem, hundreds of speckle fields were recorded with and without the sample, and the reconstructed image from these speckle illuminations was averaged to remove the speckle in reconstructed images. This method combines the advantages of incoherent imaging in resolution and image cleanness, with the merit of coherent imaging in complex electric-field recording and 3D imaging. Compared to the phase images obtained with plane wave illumination [Fig. 12(b)-(c)], the images reconstructed by using speckle illumination [Fig. 12(d)-(e)] verify the resolution enhancement. Quantitatively, the FWHM of the point spread function was 516 nm for plane illumination and 305 nm for the speckle illumination, which are very close to the theoretical expectations.

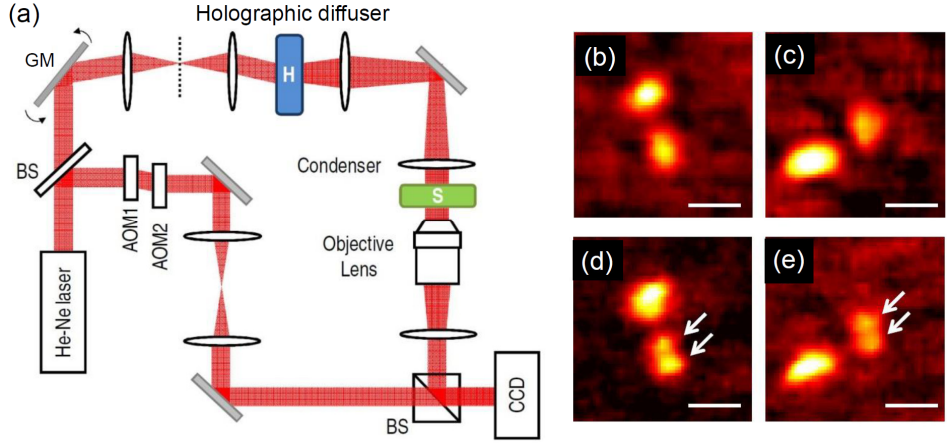

Figure 12. Schematics of DHM with sequential speckle illumination. (a) Experimental setup; (b)-(d) Phase maps of polystyrene beads (200nm diameter) under plane wave illumination (b-c) and speckle illumination(d-e). Scale bars in (d, e) are 500 nm. Images taken from Ref. [230].

It is worthy to mention that, for the averaging operation used in the above-mentioned technique, it is preferable to use a speckle field having the equal weight or power over the whole frequency spectrum. For this purpose, a highly-scattering device such as ground glasses or holographic diffusers is often used for speckle pattern generation. Instead of this, an iterative reconstruction [116] was proposed to release the requirement on the speckle illumination in QPM. Figure 13 shows the optical setup used for that study and containing a SLM for generating the speckle fields. The speckle fields illuminate the object after passing through the lens $L_1$ and the microscope objective $MO_1$. We denote with $A_o^i$ the complex amplitudes of the fields diffracted by the object illuminated with the $i$th pattern. When no object is inserted in the setup the fields corresponding to the $i$th pattern are denoted with $A_{Speckle}^i$. Holograms are recorded by superimposing a reference beam $R$ to the object $I_o^i = |A_o^i + R|^2$ and speckole fields $I_{Speckle}^i = |A_{Speckle}^i + R|^2$ from which $A_o^i$ and $A_{Speckle}^i$ can be reconstructed.

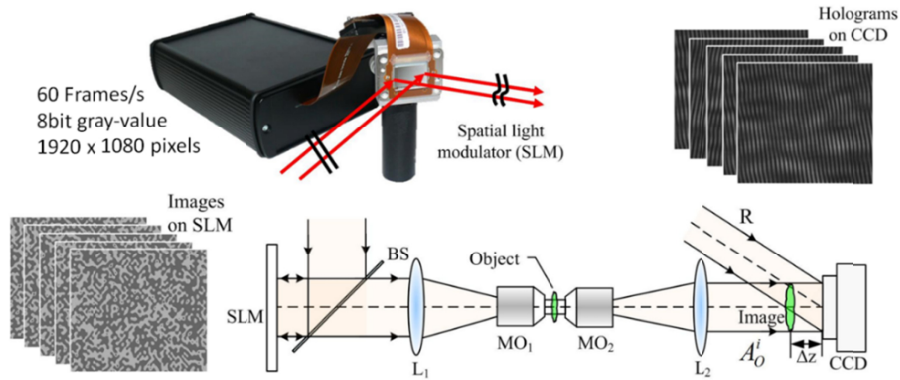

Figure 13. DHM with speckle illumination. Images taken from Ref. [116].

As previously stated, speckle fields can be understood as a combination of plane waves with various illumination directions. Each one of those oblique illumination beams will shift the spectrum of the object in the reciprocal plane [see Fig. 14(a)] allowing access to extra spatial frequencies. This extra information is integrated by an iterative method in order to synthesize a larger NA. The flowchart is shown in Fig. 14(b) and essentially involves the following steps: 1) initialize the object wave $O(x,y)$ by averaging the complex amplitude of the reconstructed object wave $O_i = 1/N \cdot \sum_{i=1}^{N} A_o^i / A_{Speckle}^i$; 2) multiply $O(x,y)$ with the speckle illumination complex amplitude $A_{Speckle}^i$ to get the complex amplitude of the speckle-based object wave $A_S^i$; 3) perform the FT of $A_S^i$, and replace its central part (circle confined to a radius equal to $NA/\lambda$), with the corresponding part of the spectrum of $A_o^i$. The replacement operation bring new high frequencies which are downshifted by the speckle illumination $A_{Speckle}^i$; 4) determine the resolution-enhanced object wave $O(x,y)$ by taking the inverse FT, and dividing the retrieved complex wave $A_{Syn}^i$ by $A_{Sto\,peckle}^i$; 5) continue the iteration by replacing the object wave $O(x,y)$ with the newly-reconstructed one. The iterative method enhances not only the resolution but also the SNR when the reconstructed object wavefronts from different speckle-based holograms are averaged. The use of an SLM for speckle field illumination avoids mechanical movements and allows high repeatability in comparison with dynamic devices.

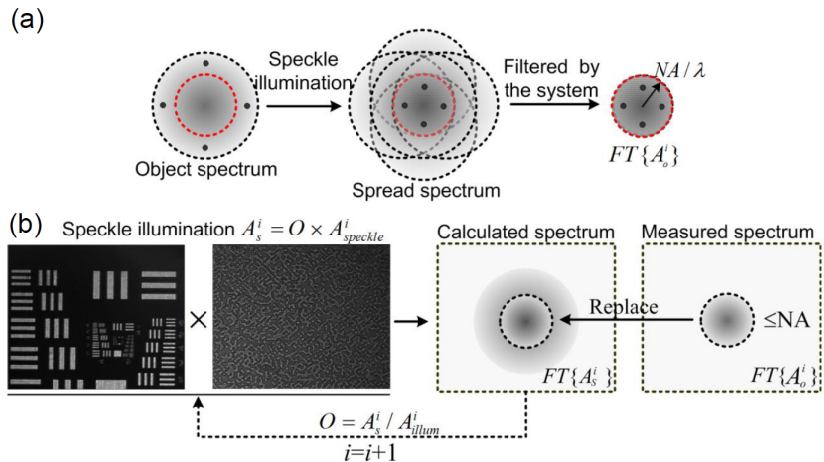

Figure 14. Iterative reconstruction of DHM with speckle illumination. (a) Frequency spectrum evolution in the imaging process; (b) iterative reconstruction with resolution enhancement. The red-dashed circles in (a) and the black-dashed circles in (b) represent the NA of the optical system. Images taken from Ref. [116].

*3.4. Resolution enhancement imaging through scattering media*

In addition to complex amplitude retrieval, holography is capable of overcoming scattering and allows imaging objects positioned behind the scattering medium via coherence gating [231-233]. The coherence gating provides the possibility sort out non-scattered (ballistic) or slightly scattered (snake-like) light to construct a high-quality image, considering the two have different optical paths through the sample. This strategy has been successfully applied using the first arriving light method for pulse lasers [234] or spatial filtering of continuous wave lasers with coherent light [235]. Other approaches yield with the possibility of imaging by means of a coherent light through a flowing turbid medium based on a Doppler frequency shift [236] and through a quasi-stationary turbid medium by multiple holographic acquisitions [237]. A combination of DH with two-point intensity correlation [238] or phase-shifting DH using temporally low-coherence light [239] to recover the phase information through a scattering medium has also been presented.

The First Arriving Light (FAL) approach was initiated as a promising direction in order to allow performing imaging via scattering media [240-243]. The FAL is propagated without scattering and thus its spatial content is not blurred, and the resolution is not lost. This information can be extracted by its proper gating from the scattered data. The gating can be done both in time as well as in the coherence domain. If a pulsed laser is used and a holographic recording is performed, the arriving light interferes with the reference beam. Both the object and the reference waves are pulsed, and the short coherence length of the light source guarantees the required time gating. Therefore, only the ballistic photons are contributing to the interference fringes of the hologram and thus only they can be decoded. Despite the fact that the reconstruction of high spatial frequencies done via FAL and time gating [244], an increase of the thickness of the scattering medium results in a situation where the SNR of the obtained FAL is too low and thus the reconstruction of the spatial frequencies of the original image becomes complicated [245].

Here, the combination of the oblique illumination and holographic recording is demonstrated for further extraction of the high spatial frequencies from the inspected object after being scattered by the turbid medium. One concept to retrieve these high spatial frequencies lies on combining a time or wavelength multiplexing oblique illumination with the FAL holographic recording. We combined the temporal gating of the FAL approach with gating in the light's angle of propagation by encoding object's angular information yields the superresolved images [245]. The angular information of the object may be encoded either by transmitting it in different wavelengths (wavelength multiplexing) or by using different time slots (time multiplexing). When time encoding is applied, each temporal slot is associated with the different angular direction (different spatial frequency). When wavelength encoding is applied, different angular directions are translated into different wavelengths while being sent at once in a single holographic recording. In both approaches, a set of holograms of the transmitted information are recorded for each angular direction, despite sparse aperture synthesis technique may be applied in order to decrease the number of required angular codes. Eventually, the image with the improved resolution is reconstructed from the digital hologram using numerical decoding.

Unlike former FAL based concepts, of which the resolution limited by the low NA of the objective, the resolution enhancement beyond NA-based resolution was achieved in Ref. [245] using a conventional coherent gating and scattering medium. The angular information of the object is encoded first. The encoding was done either by wavelength or in time domains allowing the restoration of the angular information even in presence of a highly scattering medium. The additional usage of the temporal gating reduces the requirements from the angular encoding due to its individual ability to sharpen objects spatial distribution.

Figure 15(a) presents the experimental setup used for superresolved QPM via scattering medium by using oblique illumination with wavelength multiplexing [245]. This setup is based on an interferometer that includes the reference path (the lower path) and the information path (the upper path). A Dye laser that allows a wavelength sweeping is passed through a beam splitter that initiates both paths. The illumination beam is passed through a grating that directs different wavelengths into different directions. The beam is passed through the object and through the scattering medium. A telescope system *MO-lens* with a very small aperture located in the Fourier plane of *MO* filters all the angular directions except the DC term. After the imaging system, a plane wave is obtained while its intensity corresponds to different angular information of the input object (encoded with each wavelength). Now, the beam interferes with a reference wave and the generated hologram is recorded by a CCD camera. Since the Dye laser sweeps wavelengths, each recorded hologram will correspond to a different wavelength of the input source. In Fig. 15(b) shows the images reconstructed by using a single on-axis illumination [the upper part in Fig. 15(b)] and 52 wavelength-encoded illuminations [lower part in Fig. 15(b)]. One may see that increased resolution was obtained by using the designed illumination and coherence gating. Note that also here the resolution improvement was 1D mainly along the horizontal axis.

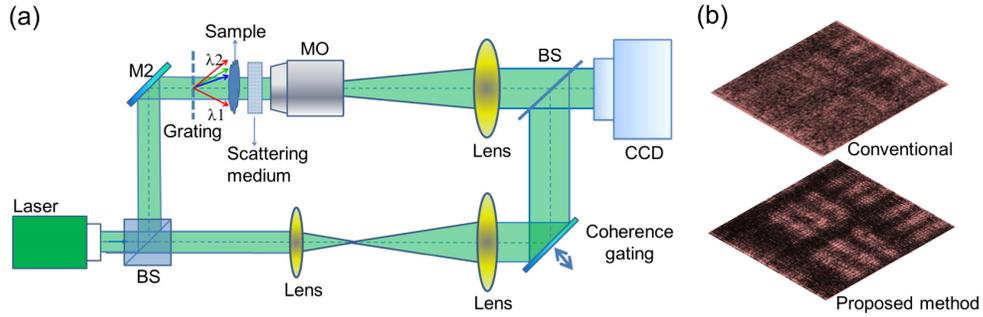

Figure 15. Superresolution QPM imaging via scattering media (a). The optical configuration used for the experiment. (b). The experimental results: the image reconstructed by using a single on-axis illumination (left) and wavelength-angular encoded illumination (right). Images adapted from Ref. [245].

Recently, a more sophisticated approach has been developed for superresolved QPM via scattering medium, which places a turbid medium between the sample and the imaging objective to improve the spatial resolution of an objective lens beyond its diffraction limit and extends its FOV [246]. This method characterizes the turbid medium with a transmission matrix, which makes the deterministic connection between the input and output of the disordered medium. The transmission matrix weights the spatial frequency of the object wave by correlating the recorded intensity with the transmission matrix of the turbid medium for each illumination angle. In practice, the transmission matrix is obtained by illuminating the medium with a laser beam at different angles ($\theta_x$, $\theta_y$), and the complex amplitude $E_{trans}$(x, y, $\theta_x$, $\theta_y$) of the generated beam is recovered by interfering with an independent reference wave [247]. This process is repeated, and 20 000 transmission fields covering the angular range of illumination corresponding to 0.5 NA were recorded. Once this transmission matrix has been determined, the turbid_medium is analogous to an unconventional lens with interesting properties. Now, the turbid medium can be used as a lens (namely turbid lens imaging, TLI) to image a sample located at the sample plane. The object is seen as a set of angular waves with their characteristic frequencies ($\theta_x$, $\theta_y$). After passing through the scattering medium, each angular wave is distorted in its own way independent of the others and is linearly superposed with the others to form a distorted image of the object $E_d(x, y)$. The weight of each angular wave can be obtained by correlating $E_d(x, y)$ with the transmission matrix $E_{trans}$(x, y, $\theta_x$, $\theta_y$):

$$A(\theta_x, \theta_y) = \sum_{x,y} E_{trans}^*(x, y, \theta_x, \theta_y) \cdot E_d(x, y) \tag{7}$$

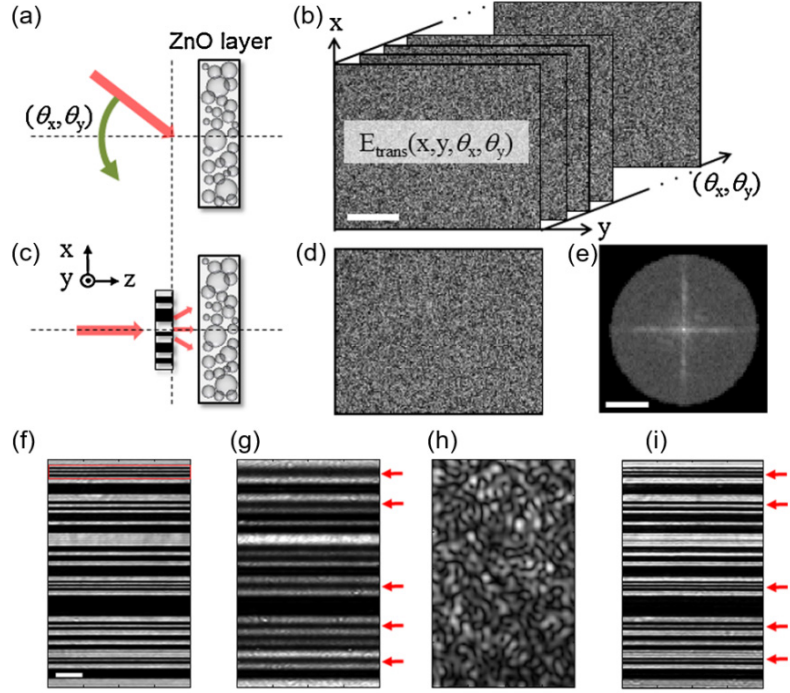

Figure 16. Schematics of TLI. (a) Recording of the transmission matrix for a disordered medium. The incident angle of a plane wave, ($\theta_x$, $\theta_y$), is scanned, and the transmitted wave is recorded at each incident angle. (b) The recorded amplitude images for different angular spectrum. (c) TLI of an object. (d) The distorted image of an object, namely a resolution target pattern. (e) Angular spectrum of the object acquired by projection operation. (f) A conventional image of the barcode-like target by a high-NA objective lens (1.0 NA) and (g) by a low-NA objective lens (0.15 NA), respectively. The fine pitches indicated by red arrows are not visible in (g). Scale bar: 10 μm. (h) A distorted image of the object taken by a low NA objective lens after inserting ZnO layers (T= 6%). (i) Object image recovered from the distorted image in (h) by TLI. Fine pitches are visible even if the image is taken by a 0.15 NA objective lens. Images taken from Ref. [246].

After the frequency spectrum of the object is determined, the complex amplitude of the object can be obtained by a reverse FT on $A(\theta_x, \theta_y)$. The spatial resolution of the reconstructed image depends on the angular spectrum coverage of the recorded transmission matrix. To demonstrate spatial resolution enhancement using a disordered medium, a barcode-like target was imaged at first by a high-NA objective lens (1.0 NA) [Fig. 16(f)], then a low NA objective lens (0.15 NA) [Fig. 16(g)], and last a turbid medium with ZnO nanoparticle layers were placed between the low-NA lens and the object [Fig. 16(i)]. It is found that the finest lines bounded by the red box in Fig. 16(f) are well resolved because their spatial period, 2.5 μm, is larger than the diffraction limit (0.77 μm). The finest pattern in Fig. 16(g) is indistinguishable due to insufficient resolving power. The finest lines are well distinguished in Fig. 16(i), implying a substantial resolution enhancement when compared with Fig. 16(g). Actually, the object spectrum from the distorted image is extended up to 0.85 NA, considering the transmission matrix was recorded by scanning the angle of the input wave from -53° to 53° along the direction orthogonal to the barcode-like lines in the object. Furthermore, the TLI also enlarges the FOV. Recently, the TLI has also been demonstrated for its application in bio-photonics [248]. A single multimode optical fiber, as a turbid medium, was demonstrated for scanner-free and wide-field endoscopic imaging [249]. TLI approach enables to retrieve the complex field maps of an input field, and it can provide not only the intensity image but also the phase image of semi-transparent samples. Based on the complex field retrieval, TLI was proved to be able to refocus a sample without physical depth scanning [249]. The TLI approach requires hundreds and thousands of illuminations to recover the image of a sample and therefore its current application is limited to static samples. This limitation will be overcome sooner or later with the advance of fast modulation and imaging devices.

## 4. RESOLUTION ENHANCEMENT IN REFERENCE-LESS QPM

Interferometric phase imaging is the most commonly used approach to retrieve phase information with high accuracy. However, sometimes the need to introduce an additional reference wave increases its complexity. Moreover, the measurements are sensitive to environmental disturbances coming from thermal and/or vibration conditions that are different on both interferometric beams. To circumvent this difficulty, common-path interferometers and reference-less phase imaging approaches have been exploited. In common-path interferometers [31, 250], both and object and reference waves pass through the same optical components to an imaging device, and no phase difference can be generated between the two waves upon environmental disturbances. Therefore, common-path QPM often has a rather high path-length sensitivity (PLS), for example a PLS of $\lambda/5500$ was reported [31], which enables to quantify the nanoscale thermal fluctuations of red blood cells (RBCs) [63]. The investigation revealed that the tension coefficient of RBCs is around $(4.1 \pm 1.1) \times 10^{-6}$ $J/m^2$, 4–24 times larger than that of vesicles. Such quantitative comparison suggests that the contribution of the cytoskeleton might be responsible for this enhancement.

Reference-less phase imaging approaches includes, just to cite a few, wavefront sensing based on Shack–Hartmann sensors [42], pyramid sensors [43], or shearing systems [251]. Other approaches are based on common-path interferometric configurations and are capable of reference beam synthesis from the object beam in order to avoid the external reinsertion of a reference beam [252-254]. Others utilize a transport of intensity equation algorithm with slightly defocused images for QPI [255-258] while others are based on the recording of different defocused images of the sample to iteratively retrieve phase information [259, 260]. Alternatively and in LHM, beam-propagation-based methods can also retrieve quantitative phase information from the recording of a series of diffraction patterns, which are usually recorded at different planes [44, 45], with different wavelengths [34, 38-40, 261], by moving sub-apertures over the sample plane [46], by modulating the object wave with different phase patterns [47, 54], by using modulated illuminations [48], illuminating the object at different angles [53], or varying the distance between illumination function and object function [55]. In addition, on-chip microscopy with geometrical SR [262-276] was also an important type of reference-less QPM.

### 4.1. Resolution enhancement approaches in LHM

LHM derives from a digital implementation of the Gabor's invention [277] where a point source of coherent light illuminates the sample and the diffracted wavefront is recorded by a digital sensor [278]. However, different with the conventional Gabor's invention, LHM uses a solid-state image recording device (typically a CCD or CMOS camera) to record the hologram instead of the holographic plate recording media. Nowadays, LHM becomes an emerging imaging modality with widespread interest where imaging is performed using a lensfree configuration.

LHM is typically implemented using two opposite layouts [262]. In the first layout, the sample is closely placed to the illumination point source and farther in comparison with the digital sensor [278-281]. And in the second layout, the illumination source is faraway while the sample is on top of the digital sensor [282-285]. The first configuration introduces a magnification factor (typically ranging from 5X to 20X) by the geometrical projection of the sample's diffraction pattern at the digital sensor plane, and provides similar FOV and resolution limit as reported in lens-based DHM with a medium NA (e.g., in the range of 0.4-0.5). The second layout provides no geometric magnification (approximately 1X range) but an extremely improved FOV since the whole sensitive area of the detector is available. This second configuration provides a modest resolution limit incoming from lower NA values (0.2-0.3 NA range), mainly because of the geometrical constraints imposed by the detector. For the sake of simplicity, let us call the first implementation as digital in-line holographic microscopy (DIHM) while the second one will be referred to as on-chip microscopy.

Although there are some approaches mainly in DIHM that externally reintroduces a reference beam for holographic recording at the output plane, we have decided to keep all those approaches in this section even in the case we are dealing with reference-less QPI. The reason is to have a global view of SR approaches in LHM instead of splitting some of them in section 3 while the majority is included in section 4 owing to the holographic nature of the Gabor in-line hologram without any external reference beam.

### 4.1.1. Digital in-line holographic microscopy with SA

In order to improve general capabilities (such as resolution in particular) of both previously commented configurations, many approaches have been reported mainly in the last decade. In the first case (sample magnified in the 5X-20X range), the resolution limit is ruled by diffraction in a similar way to that in DHM: there is an aperture in the system that limits the maximum range of spatial frequencies passing through the system. In DIHM, the NA of the imaging system is defined by the ratio $L/D$ with $L$ and $D$ being the half size of the digital sensor and the distance between the sample and the sensor. Thus, the resolution limit in DIHM can be improved by scanning a larger hologram by applying different approaches. By scanning we mean a relative motion between the object and the digital camera and it implies the generation of an expanded hologram containing a bigger portion of the diffracted object wavefront compared to the non-scanning case.

The generation of the expanded hologram can be implemented using, in essence, 4 different strategies. The first strategy (and maybe the most appealing one) deals with the displacement of the camera at the recording plane [99, 156, 158-160, 163, 286-292] thus synthesizing a larger hologram in comparison with that one provided by only one camera position. The second strategy performs SA generation by angular multiplexing the spectral object information [155, 164, 174, 181, 293-297]. Angular multiplexing is implemented by oblique beam illumination over the object in a similar way as in DHM and allows the recovery of additional spatial frequencies falling outside the digital camera sensible area when on-axis illumination is used. The third strategy is related to the use of additional optical elements (typically diffraction gratings) that are placed between the object and the digital camera [100, 298-303]. The diffraction grating directs towards the camera an additional portion of the diffracted object wavefront allowing the generation of the SA hologram. And finally, the fourth strategy is based on object movement instead of shifting the digital camera for aperture synthesis [98, 304-306].

As one can notice, the four strategies are equivalent and the effective outcome of each one of the four encoding strategies is the same one: to downshift extra content of the object spectrum that usually falls outside the limited system aperture with conventional illumination. This downshift allows the recovery of such extra apertures using different decoding methods and depending on the used encoding one. Figure 17 includes 4 cases of experimental results regarding synthetic synthesis in DIHM when considering the 4 previously reported strategies. By rows and from up to down, Fig. 17 includes (a) camera shifting, (b) oblique beam illumination, (c) diffraction grating insertion and (d) object scanning for SR in DIHM. And by columns and from left to right, Fig. 17 presents the conventional low-resolution image (left), the generated SA (center) coming from the addition of elementary apertures (white dashed rectangles) in comparison with the conventional aperture (central blue solid line rectangle), and the superresolved image as inverse FT of the SA (right). Notice that in DIHM the conventional aperture shape is rectangular rather than circular as in DHM since it is the CCD the element defining the system's aperture.

Figure 17(a) (the first row) shows the experimental results obtained by shifting the camera [289] doubling the resolution of the conventional imaging system. The camera was shifted at the image plane in the horizontal and vertical directions since most relevant information of the object is in those directions. But the SA shape can be customized according to such a priori object information. As an example, a cross-shape SA (central image) was generated with expanded frequency coverage that finally yields in an isotropic resolution-enhanced image (right image) in comparison with the conventional one (left image). The resolution limit is expanded from 12.4 μm (Group 6-Element 2 of the USAF test) until 6.9 μm (Group 7-Element 2) for the horizontal CCD direction, revealing a resolution gain factor close to 2.

Oblique beam illumination downshifts the spatial frequency information of the object so the synthetic enlargement of the system's aperture can be achieved without camera displacements. Here, an illumination pinhole is shifted to different off-axis positions in order to provide oblique beam illumination. To allow this, the sample should be static between the different recordings, so the technique is using time multiplexing for aperture synthesis. Figure 17(b) (second row) includes the experimental results reported in Ref. [295] and concerning a fixed swine sperm bio-sample. In this case, 4 apertures in the vertical and horizontal directions and 4 apertures in the oblique directions were considered. As a consequence, full 2D coverage at the Fourier domain is achieved when generating the SA. Using this strategy, a synthetic NA close to 0.7 was experimentally validated for the largest CCD direction.

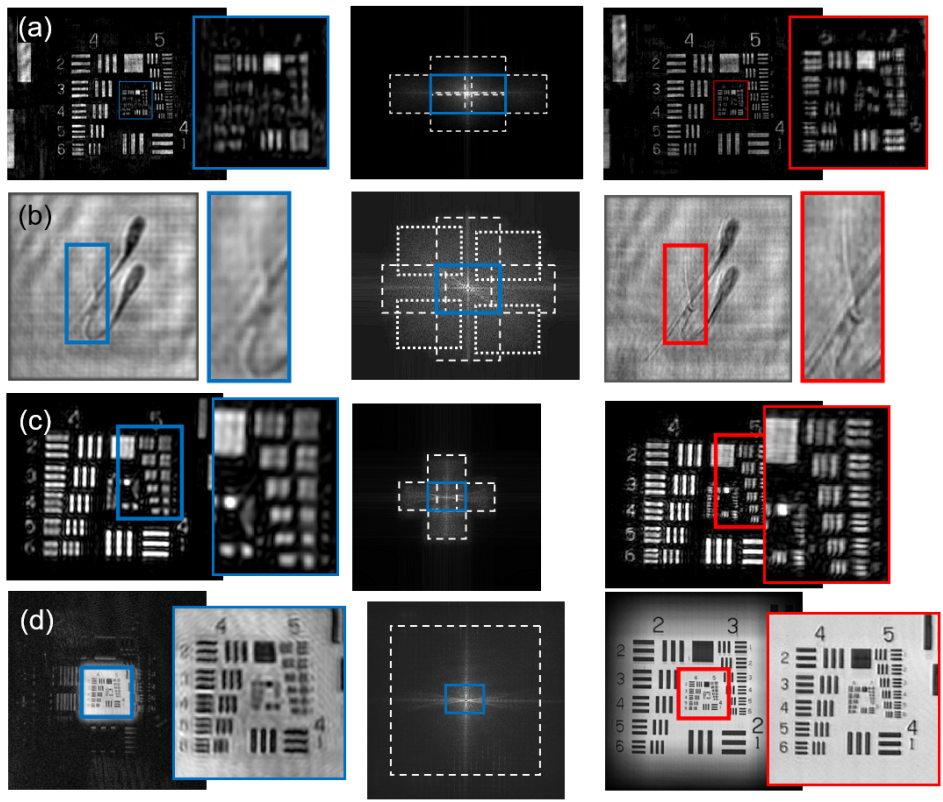

Figure 17. Experimental results for resolution enhancement of DIHM via (a) shifting the camera at the recording plane (first row – images taken from Ref. [289]), (b) using oblique beam illumination (second row – images taken from Ref. [295] – "*V. Micó and Z. Zalevsky "Superresolved digital in-line holographic microscopy for high-resolution lensless biological imaging," J. of Biomedical Optics, 15(4), 2010*), (c) inserting a grating in the experimental layout (third row – images taken from Ref. [164]), and (d) considering object movement at the input plane (fourth row – images taken from Ref. [305]). Inside each column, the conventional

image without aperture synthesis (left images), the generated SA (central images), and the finally retrieved superresolved image (right images). In each row, the central parts were magnified to allow easy identification of the SR effect.

Figure 17(c) (the third row) depicts the experimental results concerning SR imaging by properly inserting a diffraction grating in the lensless Fourier holographic experimental layout [300]. Depending on the Fourier spectrum of the diffraction grating, the SR effect can be 1D or 2D. Two possibilities can also be considered for retrieving the complex amplitude distribution of each band-pass image depending on the mode that the reference beam is introduced in the recording plane: on-axis recording with PS algorithm and off-axis recording with Fourier domain filtering. Thus, it is possible to choose between a superresolved image over a large object FOV or a higher number of band-pass by limiting/masking the object´s FOV. Here, a cross-shape SA is generated yielding in a resolution gain factor of 2 in both directions. According to the images included in Fig. 17(c), the resolution limit is improved from 31.25 µm (Group 5-Element 1 of the USAF test) until 15.6 µm (Group 6-Element 1) for the horizontal CCD direction.

Last but not the least, aperture synthesis has also been reported by using the object translation as a key tool for image quality improvement in LHM [98, 305]. Lensless Object Scanning Holography (LOSH), profits of the linear movement of the object to record a set of reflective digital lensless Fourier holograms. Because of the linear object displacement, different spatial information coming from different object's regions of interest (ROIs) is recorded in each hologram and retrieved later on. Meanwhile, every single ROI passes along the whole FOV in front of the CCD and becomes illuminated with a different oblique beam for every position, thus allowing for retrieving different spatial frequency content of every ROI. Those two characteristics produce object FOV enlargement and resolution improvement, respectively, when considering the whole set of recorded holograms. Moreover, the SNR of the final image is also enhanced due to the averaging of the coherent noise when adding the whole set of recorded holograms. As a result, the final image provided by LOSH resembles an image obtained under white light illumination but considering that not only intensity but also phase information is accessible. Thus, an extended DOF image can be also synthesized owing to the coherent nature of LOSH by numerical propagation to different axial distances. Figure 17(d) (last row) shows the experimental results provided by LOSH where an impressive image quality improvement concerning FOV, resolution, and SNR is achieved for a 2D resolution test. The scanning was performed using a regular raster composed of 20 rows and 20 holograms per each row. As a consequence, the conventional FOV ($1.5 \times 1.5$ mm, approx.) is expanded up to a synthetic FOV 3.7 time higher ($5.5 \times 5.5$ mm, approx.), the resolution limit is enhanced from 44 µm to 14 µm meaning a resolution gain factor equal to 3.2, approx., and the SNR value is improved from 2.34 (single hologram) to 10.16 (superresolved image) after applying LOSH yielding in a SNR gain factor of 4.34.

### 4.1.2. On-chip microscopy with geometrical SR

Coming back to our two opposed layouts in LHM, resolution enhancement methods have also been reported using on-chip microscopy (1X sample's magnification). In the on-chip microscopy, the sample is placed really close to, or, directly contact with the imaging sensor, so the limiting factor in resolution is essentially the pixel size of the digital sensor that defines the sampling requirements and the coherence properties of the illumination source which defines the maximum angle that a diffracted beam will interfere with the non-diffracted reference beam. In this configuration, an immersion medium with the refractive index higher than 1 fills the gap between the sample and the sensor since there is glass coming from the coverslips containing the sample. Such immersion scheme enhances the resolution in comparison with the previous DIHM layout where the air is filling this gap.

Nowadays, there are available sensors having a pixel size at the micron range yielding resolution limits that are similar to that of the DHM systems having microscope lenses with 0.25 NA for a visible illumination. Despite the spatial resolution can be improved by reducing the pixel dimensions, the SNR will be reduced if doing this. To circumvent this conflict between the resolution and the SNR, pixel SR (or geometrical SR) was proposed and successfully applied to on-chip microscopy [262-276]. This strategy records sequentially multiple images of the same sample when the sample is shifted along the horizontal and vertical directions for a sub-pixel distance each time. These sub-pixel shifts synthetically reduce the effective pixel size of the camera, so the final resulting image exhibits a higher resolution limit from a sampling point of view. We note that, shifting the sample with respect to the camera will not generate a SA but it improves the sampling when the pixel size of CCD pixels is fixed. In aside of relative shifting between the sample and the camera, relative shifting the sample and patterned illumination [307] or introducing a numerical lens in the convolution based reconstruction algorithms can also results in a reduced virtually pixel size [308].

Figure 18 includes some representative images from these types of approaches [272]. The experimental layout [Fig. 18(a)] consists on a spatially incoherent light source attached to an optical fiber (~50 μm core diameter acting as illumination pinhole) which is shifted to several (up to 36) 2D scanning positions depending on the pursued resolution gain factor. The layout distances represented in Fig. 18(a) are $z_1$ ~10 cm and $z_2$ ~ 0.75 mm, and the CMOS sensor (2.2 μm × 2.2 μm pixel size) provides a total FOV area of ~24.4 mm$^2$. With this distances, the shift at the sensor plane ($s_{sensor}$) can be computed from the shift of the source ($s_{source}$) as $s_{sensor}$ ~ $s_{source}$ ($z_2$ $n_1$) / ($z_1$ $n_2$), where $n_1$ and $n_2$ are the refractive index values of the spaces source-sample ($n_1 = n_{air} = 1$) and sample-sensor ($n_2 = n_{glass} = 1.5$). Using the previous distances results in $s_{sensor}$ ~ $s_{source}$/200 meaning that subpixel shifts can be easily implemented from larger source displacement (~0.1 mm range). Under this configuration, the authors demonstrated SR imaging (corresponding to a factor of 6) in both amplitude and phase distributions of a wide variety of samples: a specially fabricated grating (2 μm pitch), red blood cells coming from a blood smear sample and Caenorhabditis elegans bacterium. The final images [see Fig. 18(b)-(c)] achieve ~0.6 μm spatial resolution coming from a ~0.5 NA over the whole sensor FOV.

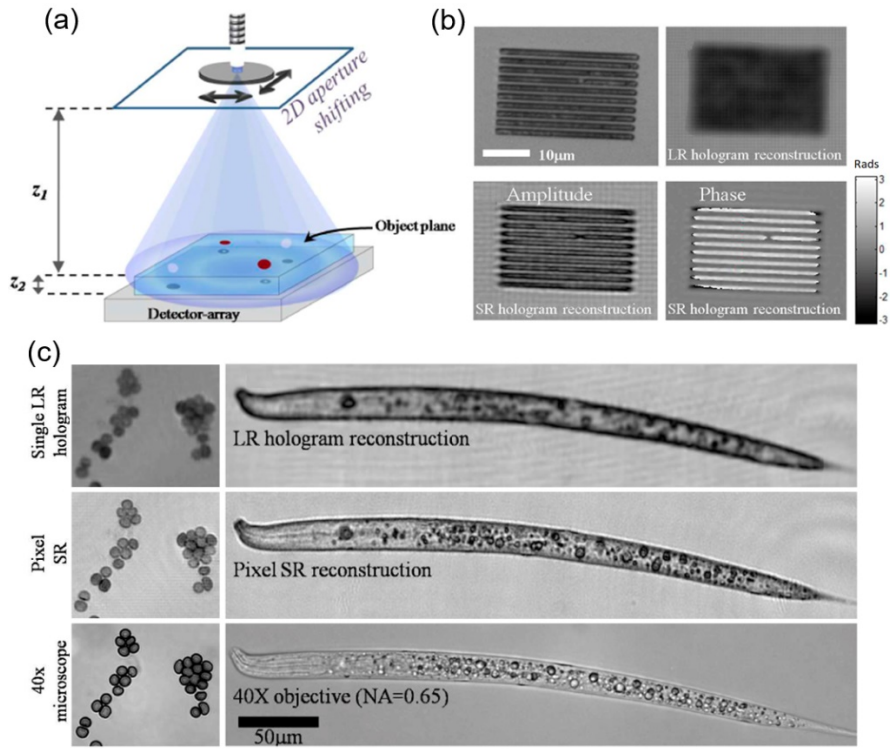

Figure 18. Lens-free on-chip microscopy using pixel SR approach over a wide field-of-view: (a) scheme of the experimental layout, (b) experimental results when a phase grating is used as calibration object, and (c) results of a RBCs cluster (left column) and Caenorhabditis elegans (right column). The comparison was done for the conventional low-resolution on-chip imaging case with the superresolved images in amplitude and phase. Images adapted from Ref. [272]. LR in (c) indicates low-resolution.

Similar to the previous on-chip microscopy configuration, the shift is relative between the sample and the digital recording device. So, it is possible to shift the digital sensor [275, 276], the sample [273, 274], or the illumination [269-272] but keeping static the rest of the setup. Nevertheless, the most practical technique is to shift the illumination source because precise subpixel shifts can be generated from less-precise illumination shifts due to the geometrical characteristics of the layout (distance's ratio between the illumination-to-sample and the sample-to-digital sensor). Nevertheless, other approaches have been reported using wavelength scanning [267, 268] or multiple intensity images at different sample's distances [265, 266]. Using these types of approaches, the resolution limit has approached the diffraction limited resolution limit with half-pitch resolutions equivalent to those from 0.8-0.9 NA microscope lenses [262, 269].

Finally, similar to the previous DIHM configurations, SA generation has also been proposed as a way to increase the resolution limit in on-chip microscopy [264]. Here, the illumination source is not only shifted to achieve geometrical SR by subpixel shifts but tilted to allow oblique illumination and aperture synthesis. Thus, a set of multiple low-resolution holograms (typically between 16 and 64) for each tilted beam illumination position (11 oblique illuminations for each orthogonal direction) are recorded and then used to achieve geometrical as well as diffraction limited superresolved imaging. By digitally combining all those lens-free holographic measurements, the authors clearly resolved 250 nm grating lines under 700 nm illumination wavelength which effectively corresponds to 1.4 NA.

In addition to those holographic approaches, on-chip microscopy with enhanced resolution has also been applied to other imaging modalities such as fluorescent imaging [309-312], SI [309], holographic color imaging [267, 270, 313] and as specific arrangement for cell culture [314, 315] and waterborne parasites analysis [283] in combination with deep learning microscopy [316, 317] as well as 3D tracking of sperm cells trajectories [318-320] based on improved axial resolution [321]. Notably, sub-pixel movement between the sample and the camera has achieved modest improvement in lateral resolution in vertical scanning interferometer, exceeding the Nyquist–Shannon limit of the sensor [322].

*4.2. Spatially-multiplexed interferometric microscopy with aperture synthesis*

Reference-less QPM can be realized by generating an effective reference wave from the object wave [323]. One of the most representative of this type of approaches is point-diffraction phase microscopy [28, 31, 39] which splits the object wave into two identical copies following a common path and generates a reference beam by pinhole-filtering on one those replicas. This configuration has the advantages of being less sensitive to environmental vibration thus providing high lateral and phase resolution. Aimed in this direction, there are nowadays different configurations for achieving QPM using common-path layouts [179, 251-254, 324-335], some of them implemented in regular white light microscopes with the added value of updating a regular microscope with coherent sensing capabilities [179, 251, 254, 324, 328-334].

Among those techniques, SMIM (initials incoming from Spatially-Multiplexed Interferometric Microscopy) proposes a simple and low-cost way to convert a commercially available regular microscope into a holographic one with only minimal modifications [179, 332-334]. SMIM implements a common-path interferometric setup [Fig. 19(a)] into a real microscope embodiment where: i) the broadband light source is replaced by a coherent one; ii) a 1D diffraction grating is properly inserted at the microscope embodiment; and iii) a clear region at the input plane is saved for reference beam transmission. Using SMIM, a regular microscope is converted into a holographic one working under off-axis holographic recording with Fourier filtering [332] or slightly off-axis recording with phase-shifting algorithm [333]. The appealing advantage above the conventional DHM is the stability (less sensitive to environmental disturbance) due to the common-path configuration.

The main drawback of SMIM is the FOV restriction imposed by the need to transmit a clear transparent reference beam for the holographic recording. Despite the FOV is penalized, the resolution will be improved in SMIM without modifying a given optical imaging configuration. In order to improve the resolution of SMIM, oblique beam illumination has been recently adapted to the SMIM architecture [179]. Oblique beam illumination is achieved by lateral displacement of the coherent source to a set of off-axis positions, thus allowing a complementary spatial-frequency content diffracted on-axis for each considered oblique beam (a regular procedure in oblique illumination techniques). And SMIM technique recovers the complementary spatial-frequency content provided by each oblique illumination. In addition to the on-axis one, the full set of oblique beam illuminations are used to expand up the system's aperture by generating a SA having a cutoff frequency two times higher than that of the conventional one under on-axis illumination. All together, superresolved SMIM adapts a commercially available non-holographic microscope into a superresolved holographic one. When applying SR to SMIM with the 5X/0.15NA lens, the resolution is doubled and the useful FOV of is reduced to one half [333]. The FOV is still equal to the one provided by the 10X/0.30NA lens which is often employed to reach the same resolution with superresolved SMIM. Figure 19(b)-(d) depicts the experimental results provided by this method [179] when using a 5X/0.15NA objective lens. The conventional image [Fig. 19(b)] corresponding with on-axis illumination mode is restricted in resolution and comes from the inverse FT of the conventional pupil (blue circle in the central image). Then, an oblique beam illumination procedure is implemented allowing the recovery of 4 extra off-axis apertures that are used to synthesize the expanded aperture [Fig. 19(c)]). Finally, the superresolved image [Fig. 19(d)]) showing a resolution gain factor of 2 is achieved from the SA.

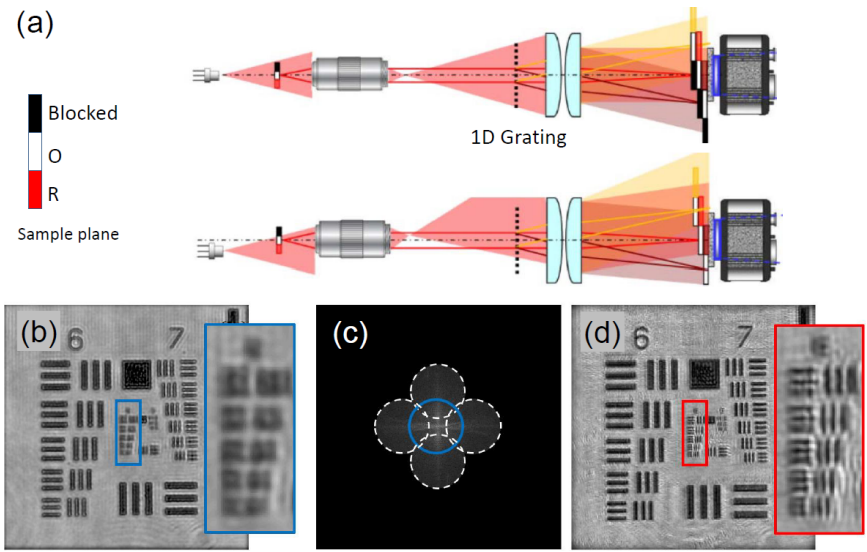

Figure 19. Superresolved imaging of SMIM using oblique beam illumination. The configuration of SMIM using oblique beam illumination (a), the low-resolution image retrieved with on-axis illumination (b), the generated SA (c), and the finally retrieved superresolved image (d). Images taken from Ref. [179].

*4.3. Reference-less QPM with oblique illumination*

One interesting approach for resolution-enhanced reference-less QPM comes from the combination of oblique beam illumination and phase retrieval using an iterative algorithm [188, 336]. The basic idea of the iterative algorithm is to record a sequence of defocused intensity images at $N$ different planes, and then process these images to extract the phase encoded in the defocused intensities (Fig. 20). The defocused intensities are generated by linear translation of the camera [259] at different axial direction or using a tunable lens [260]. The procedure is quite similar to the previous references [259, 260, 337] but now oblique beam illumination is considered aside of the on-axis conventional illumination. Thus, the phases at multiple illumination angles are retrieved and a SA can be generated by the coherent addition of the retrieved information coming from different shifted pupils. The main drawbacks of the method are: i) the sequential recording of the different $N$ intensity images that prevents the usage of the technique for fast dynamic samples; ii) the need for mechanical scanning of both the oblique beam illuminations (using galvo mirrors) and the camera shift (linear motorized translation stage), causing potential drifts, make the SA difficult.

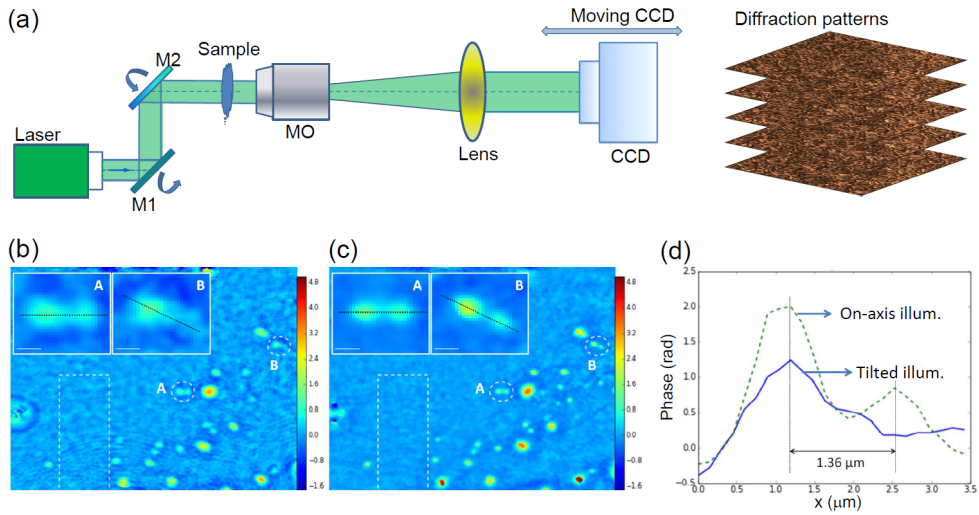

Figure 20. Iterative phase retrieval with oblique illumination. (a) Experimental setup and the schematic diffraction patterns recorded from different axial planes by moving the CCD camera for one illumination angle (b) reconstructed phase image with on-axis illumination and synthesized image with multiple oblique illuminations. (d) comparison of the line profile crossing two beads indicated with B. Images taken from Ref. [188].

Another resolution-enhanced reference-less QPM is based on phase contrast imaging, which can convert the phase distribution of a transparent object into intensity modulation by introducing a phase shift between the non-diffracted and diffracted components of the object wave in Fourier plane [31, 198, 338-340]. When this method is combined with oblique illuminations, phase imaging with higher lateral resolution and low coherent noise can be achieved [198]. The schematic setup for the quantitative phase contrast microscopy with oblique illumination is sketched in Fig. 21. 24 points periodically located on a ring were used as illumination sources, which are incoherent with each other. The light from these pint-alike sources, after being collimated by the condenser, illuminate the sample positioned in the back focal plane of the objective along different directions. The specimen is imaged by a telescope system comprised of the objective lens and a tube lens to a CCD camera. To perform phase contrast imaging, a phase plate is placed in the back focal plane of the objective, and is used to shift the phase of the undiffracted component of the object waves. The phase plate is etched in a plate made of silica. For each illumination point on the amplitude mask, three steps, corresponding to the phase 0, $-\pi/2$, and $\pi/2$, are alternately distributed in each azimuthal period of 15°, as shown in Fig. 21(a). By rotating the phase plate, different phase steps on the ring can be chosen to cover and thus delay the undiffracted components of the object waves, providing three phase-shifted interferograms. By using the standard reconstruction algorithm [198], the phase of the object can be evaluated quantitatively, and Fig. 21(b) shows a representative phase image of a tested micro-lens array.

With this method, the imaging resolution is enhanced by the oblique illumination (0.15 $NA_{\text{illum}}$). To prove this, an amplitude grating with period 2 μm and duty cycle 50% was used as a specimen, Fig. 21(c) shows the recorded the images under on-axis illumination (top) and the multi-direction off-axis illumination (bottom). The structure of the grating cannot be resolved in the image obtained with the on-axis illumination. On the contrary, the structure of the grating is clearly resolved in the image obtained by using the off-axis illumination. Besides, the visibility of coherent noise is reduced to $1/\sqrt{24}$ considering the visibility of the coherent noise decreases with the number of independent light sources in the relation of $V \propto 1/\sqrt{N}$. Furthermore, spatial light interference microscopy (SLIM) [341-343] utilizes a similar concept and achieves similar advantages. Specifically, SLIM uses a ring-shaped light source to illuminate the sample and uses accordingly ring-shaped phase shifters displayed on a SLM to retarder the non-diffracted components of the object waves for different phase values. Therefore, quantitative phase imaging can be performed by SLIM with considerably low coherent noise and high spatial resolution. Eventually, it is also worthy to point out that, the resolution enhancement of the phase contrast imaging with tiled illumination is different with the other QPM approaches with coherent illumination. This is because here the apparent improvement in the resolution is mainly due to an incoherent illumination is used, instead of the coherent illumination. Therefore, the resolution is improved slightly from $0.82\lambda/NA$ to $0.61\lambda/NA$.

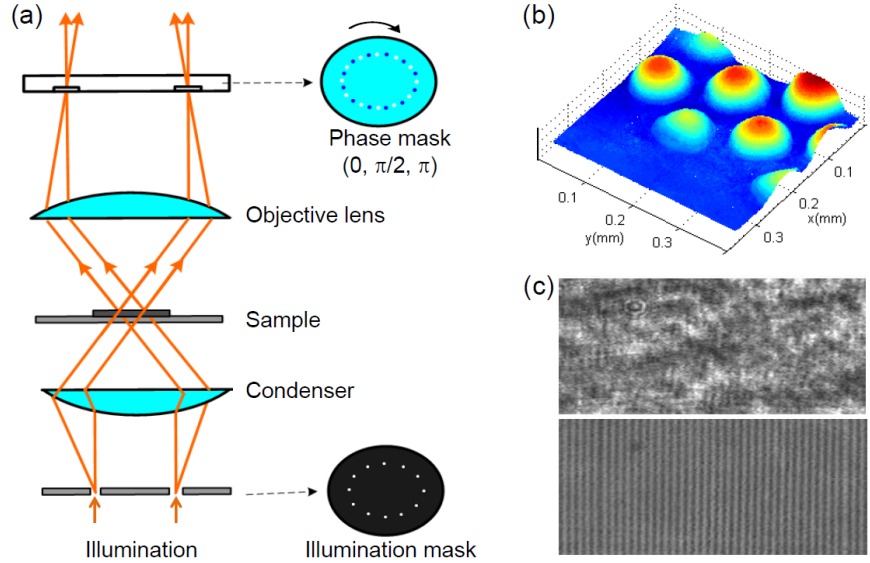

Figure 21. Quantitative phase contrast microscopy with oblique illumination. (a) Schematic configuration. (b) Representative phase image of micro-lenses fabricated on a glassplate. (c) Resolution test with a grating-like sample. The top and bottom are the images obtained with conventional on-axis illumination and oblique illumination. Images taken from Ref. [198].

### *4.4. Reference-less QPM with modulated illumination*

Beam-propagation based method retrieves the phase of a beam by iterative propagating of the wave among a sequence of diffraction patterns. The diffraction patterns may be recorded at different axial planes, with different wavelengths, by flipping the sample, or by scanning an aperture over the object wave. Partially developed speckle field was used to enhance the convergence of beam propagation-based phase retrieval [224]. The phase can also be estimated by modulating the object wave with different phase patterns [47, 54]. The third single-beam QPM with resolution enhancement lies on modulated illumination [48, 344]. Figure 22(a) shows the setup used for single-beam QPM which is similar to the one used in Section 3.4 but no reference wave is needed. The random patterns on SLM [insets in Fig. 22(a)] are projected to the sample plane by the relaying system ($L_1$-$MO_1$). After passing through the sample placed in the focal plane of $MO_2$, the object wave is imaged by the system $MO_2$-$L_2$ and the diffraction pattern [Fig. 22(a), inset] are recorded by a CCD camera located at a distance $\Delta z$ from the image plane IP of the specimen. Because these random phase patterns are generated using an SLM, they are highly repeatable and their complex amplitudes can be measured in advance. We denoted them with $A_{illum}^k$, and $k=1,2,\ldots M$ while $I^k$ denotes the diffraction pattern of the object wave under $A_{illum}^k$. The phase retrieval is performed using the following steps: (1) multiply the amplitude of the k$^{th}$ diffraction pattern with a random initial phase factor exp(i$\varphi^k$); (2) propagate $\sqrt{I^k}$ exp(i$\varphi^k$) to the object plane; (3) the calculated wave is divided by the k$^{th}$ illumination amplitude $A_{illum}^k$, and multiplied by the (k+1)$^{th}$ illumination amplitude $A_{illum}^{k+1}$; (4) propagate the newly-obtained object wave to the CCD plane; (5) replace the amplitude of the obtained object wave with $\sqrt{I^{k+1}}$; (6) repeat the iteration loop (2)–(5) by using k+1 instead of k, until the difference between two neighboring reconstructions will be smaller than a threshold value. Furthermore, the object waves reconstructed from different groups of those random-phase illuminations are averaged in order to reduce the noise. Figure 22(b) shows a representative phase image obtained by using single-beam QPI. Notably, the spatially modulated illumination also improves the spatial resolution of the phase images since it scrambles the specimen spectrum and, thus, some high-frequency components of the object wave beyond the NA of the system are downshifted and can pass through the limited system aperture. These high-frequency components are shifted back to their original positions during the iterative process. Figure 22(c) shows the resolution enhancement of single-beam QPM compared with the DHM with plane wave illumination.

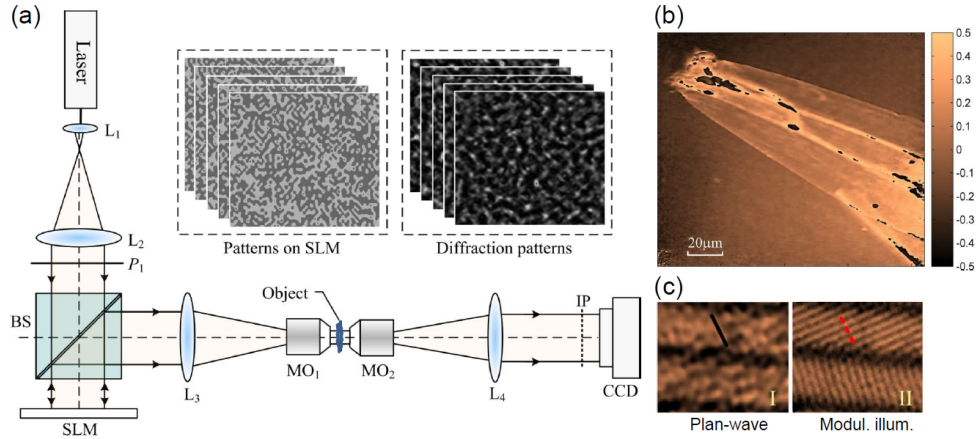

Figure 22. Single-beam QPM using modulated illumination. (a) schematic setup; insets in (a) show the phase patterns displayed on SLM (left) and the diffraction patterns recorded by the CCD camera (right); (b) Representative phase image obtained by using single-beam QPM; (c) resolution comparison of DHM with plane wave illumination and single-beam QPI with modulated illumination. Images taken from Refs. [48, 344].

Recently, random binary amplitude masks, generated by a digital micro-mirror device, were proved to achieve a real-time wavefront acquisition based on an iterative phase retrieval algorithm [345]. Instead of recording the Fresnel diffraction patterns under random-pattern illuminations, the object's complex amplitude can be obtained from the irradiance distribution measured in the Fourier plane of the lens, when a sequence of illumination amplitude patterns and a single focusing lens were used [346]. The same setup can also be used to demonstrate a new adaptive iterative phase retrieval method, where the negative phase of the object (phase edge) in the intermediate image plane will be loaded on SLM. If we write in the correct negative phase it is possible to eliminate the edge effect almost completely, and thereby we retrieve the phase of the edge with improved measurement accuracy [347].

### *4.5. Fourier ptychographic microscopy*

Another reference-lens QPM approach with resolution enhancement is termed with Fourier ptychographic microscopy (FPM) [53, 348-351]. The name 'Fourier ptychography' comes from ptychography [352-355], a technique firstly invented by Walter Hoppe [356], aiming to solve the phase problem with the application of certain constraints. Ptychography uses a focused beam to illuminate the sample and records multiple diffraction patterns as a function of sample positions. This set of diffraction patterns are then used to recover the complex amplitude of the sample following an iterative phase retrieval strategy [354]. By contrast, Fourier ptychography, introduces the 'angular diversity' concept to simultaneously expand the Fourier band-pass and recover the 'lost' phase information [351]. This method uses successive oblique illuminations at different directions to recover a high-resolution and high-SBP output image by iteratively stitching together a number of low-resolution images in Fourier space.

Figure 23 shows the schematic configuration of FPM, where a sample is placed at the focal plane of a microscope objective often with a low NA. The light from different LEDs in a source-coded LED array is used to illuminate the sample at different /orientations. The direction of each two neighbouring illuminations are set close enough to reach ~60% overlap between each image's spatial frequency support. A sequence of images are recorded when the sample is successively illuminated by the plane waves at different angles from different LEDs [Figs. 23(c3-c4)]. Then, an iterative process is used to reconstruct a high-resolution image from the recorded low-resolution images. (1), initialize with the high-resolution image, $\sqrt{I_h}e^{i\varphi_h}$, with $\varphi_h=0$ and $I_h$ as any upsampled low-resolution image as a starting point. (2) generate a low-resolution image $\sqrt{I_1}e^{i\varphi_1}$ by selecting a circular region on the spectrum of $\sqrt{I_h}e^{i\varphi_h}$ and taking inverse FT. The position of the circular region corresponds to a particular angle of illumination, and the size of the region given by the coherent transfer function of the objective lens. (3) replace $I_1$ by the measured intenisty $I_{1m}$, and update the corresponding region of $\sqrt{I_h}e^{i\varphi_h}$ in Fourier space. Here the subscript m indicates the m$^{th}$ measured intensity. (4) repeat steps (2)-(3) for other plane-wave incidences (total of $N$ intensity images). (5) repeat steps (2)-(4) till a self-consistent solution is achieved. Eventually, the converged solution in Fourier space is transformed to the spatial domain to recover a high-resolution image of the targeted sample with a dramatically increased SBP (namely, high-resolution and wide-FOV). The comparison between Fig. 23(c5) and Fig. 23(c2) shows clearly the resolution enhancement by the oblique illuminations.

In FPM, the amount of overlapping between the apertures/spectra for properly stitching the collected set of low-resolution images as well as the resolution reconstruction outcome are dependent on the SNR of the collected images as well as on the experimental conditions. Previous investigations were able to show that there is a strong SNR and overlap dependency on the maximum recoverable error for low SNR, but at some point the SNR dependency is lost and the maximum recoverable error becomes fixed for a particular amount of overlapping [357]. Further to that, in Ref. [358] the authors show that the reconstructed image quality is improved by at least 4% with Fermat spiral scan compared with the image obtained using the concentric scan pattern. They also show that the Fermat spiral pattern scan increases image quality and reconstruction tolerance on data with imperfections, especially for low overlapping conditions and for low SNR.

It is worthy to point out that the SBP increase is paid with long acquisition time, typically in the order of minutes. To study live samples, which is continuously evolving at various spatial and temporal scales, faster capture time is required [253, 359]. In FPM, high-speed aperture synthesis is also recently reported by using light-splitting with gratings [360], or source-coded LED array, together with an improved algorithm [359, 361]. In order to further improve the resolution enhancement range, an oil-immersion condenser was employed to enlarge the illumination angle [362], in aside of using a high NA imaging objective [350]. FPM is often applied in the transmission mode, but recently has also been extended to the reflection mode [363]. Recently, the concept of Fourier ptychography has also been applied to fluorescence imaging with resolution enhancement [364]. Fourier ptychography can also be used for information multiplexing and coherent-state decomposition [365], considering information multiplexing is important for biomedical imaging and chemical sensing. For this purpose, multiple light sources are lit up simultaneously for information multiplexing, and the incoherent mixture of FPM acquisitions can be decoupled to recover a high-resolution sample image by using the state-multiplexed Fourier ptychography recovery routine [366]. Single-shot FPM was reported to significantly improve the imaging speed by using multiplexing illuminations from lens-array [367].

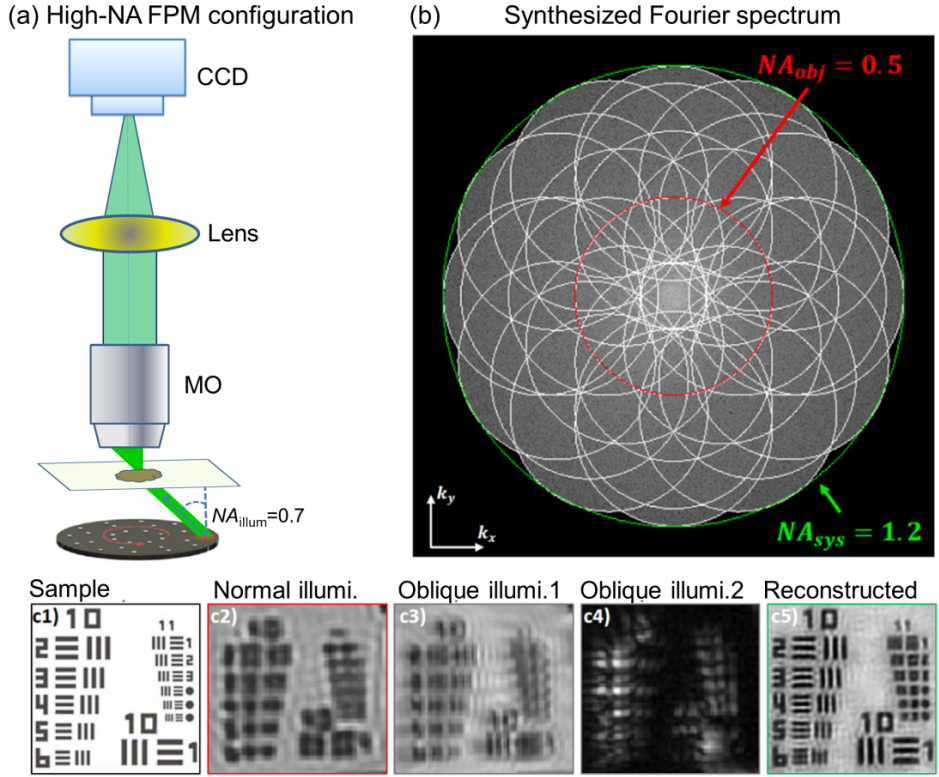

Figure 23. Principle of FPM. (a) Schematic configuration of FPM high NA, and (b) enlarged spectrum by using different illumination angles. Center red circle: Fourier support of the original microscope; white circle: Fourier support of one LED; green circle: synthesized Fourier support of the FPM system. (c1) Known sample intensity; (c2) image captured by a conventional 20X microscope corresponding to red circle in (b); (c3-c4) two images captured with different off-axis LEDs on, corresponding to two of the white circles; (c5) FPM reconstruction, corresponding to the green circle. Images taken from Ref. [350].

Besides the resolution-enhanced approaches by using a LED array, ptychography imaging by laterally translating samples through a fixed converging beam can also improve the resolution of phase imaging [368]. This is mainly due to the SA generated by illuminating the specimen with different phase gradient of the illumination beam. Furthermore, the subpixel shifting (between the sample and CCD camera) and intrinsic extrapolation effect of the iterative reconstruction also contribute to the resolution enhancement. Aside of ptychography, annular illumination was used to improve the spatial resolution of on Zernike phase contrast [198, 341, 343] or transport-of-intensity equation based QPM [369].

## 5. OTHER APPROACHES FOR SR IN COHERENT IMAGING SYSTEMS

*5.1. Second harmonic generation DHM*

Second harmonic generation (SHG) microscopy was first introduced in the 70s some years after the laser development [370, 371] and expanded rapidly to different modalities such as, scanning microscopy [372] and third harmonic generation microscopy [373]. Because of its capability to image certain structures such as microtubule, myosin and collagen, SHG microscopy is nowadays a useful technique in in-vivo biological studies [374, 375].

SHG can be described as a coherent nonlinear process in which two exciting photons are transformed by a given material into one photon with the double frequency or, in other words, with half of the wavelength of the excitation illumination. Since the resolution limit in coherent imaging is proportional to the wavelength used to illuminate the sample, a reduction factor of 2 in the illumination wavelength provides a resolution gain factor of 2 in comparison with the one obtained when using the excitation illumination. Thus, the spatial resolution is limited by the diffraction of light at the half wavelength and the temporal resolution is limited by the laser pulse duration. This is why SHG can be understood as a SR method for coherent imaging.

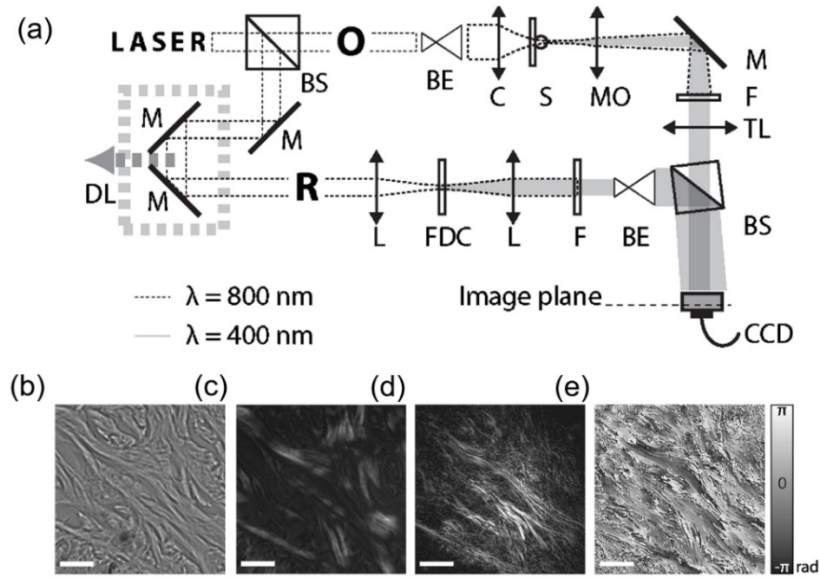

Figure 24. Label-free phase imaging of biological specimen (mouse tail dermis and epidermis) by second-harmonic DHM: (a) experimental layout working in off-axis configuration, (b) bright field image. (c) cross-polarization image showing phase retardation caused by the birefringence of collagen, (d) SHG amplitude and (e) phase (wrapped) reconstructed from a single hologram. All images present the same region of the specimen and scale bars are 10 $\mu$m. Images taken from Ref. [378].

Originally developed as a nonlinear incoherent imaging technique to provide better resolution limit with improved contrast, SHG microscopy rapidly contributed to the emergence of coherent nonlinear microscopy. This coherent nature of SHG scattering is capable of interfering with a second harmonic reference beam. The combination of SHG microscopy with DH further improves contrast imaging capability since holography retrieves complex amplitude distribution of the second harmonic generated signal from which superresolved phase-contrast imaging of the sample is possible. SHG DH was introduced one decade ago for constructing an ultrafast four-dimensional contrast microscope with high spatial and temporal resolution [89]. The technique records holograms incoming from the interference of second harmonic signal scattered from SHG nanocrystals and an independently generated second harmonic reference. It has been proved to be a powerful technique to achieve aberration-free, shot-noise-limited performance with low photon count signals, laying a sound basis for molecular biomedical imaging applications.

Since then, SHG DHM (or simply named as second harmonic phase microscopy) has been reported in numerous applications. Hsieh et al. reported on a subsequent experiment employing a Ti:sapphire oscillator for the recording of SHG holograms from HeLa cells labeled with BaTiO3 nanoparticles [88]. Here, the second harmonic is generated from the nanoparticles since the HeLa cells do not allow SHG. The same group of researchers demonstrated focusing coherent light on a nanoparticle through turbid media based on digital optical phase conjugation of the SHG field from the nanoparticle [376]. Shaffer et al. studied the second-harmonic signal generated at the glass/air interface of a microscope slide [377] as well as presented for the first time label-free holographic SHG phase contrast imaging of biological specimens [378]. Based in an off-axis DHM configuration, Shaffer et al. presented a non-scanning, single-shot image acquisition technique, limited only by the camera frame rate and using SHG signal. Figure 24 presents the experimental layout as well as quantitative phase imaging of a mouse tail dermis and epidermis achieved by the proposed SHG DHM.

Label-free SHG DHM was reported for biological samples (corn seeds, potato starch, and human muscle fibrils) with an ultrashort Yb:KGW laser oscillator by Masihzadeh et al. [379]. In that work, the authors demonstrated the capturing of a 3D image volume of a biological specimen with an exposure time of ~ 10 ms and with an SNR > 30 dB, opening the door for high-speed 4D spatio-temporal imaging in biological applications. Again, Shaffer et al. validated 3D-tracking of SHG emitters by QPI at nanometric scale [380]. Aside of on-axis [87, 379] and off-axis [376, 378-380] holographic configurations, SHG DHM has been combined with PS interferometry [381] and with SI for further resolution improvement [382]. Spatial resolution and contrast have also been reported by a method based on the intensity difference between two images obtained with circularly polarized Gaussian and doughnut-shaped beams [383].

Finally, SHG has also been merged with lensless microscopy as a way to increase the potential use and capabilities of LHM [280, 384]. This combination provides a significant reduction in the system dimension as well as improves the spatial resolution and contrast in LHM. In addition to the reduction in optical aberrations and the increase in the FOV provided by conventional LHM in comparison with regular DHM, SHG LHM increases the throughput of classical LHM since it also improves the overall photon collection efficiency as it eliminates the amount of light loss through a microscope and aperture. This fact can be of particular interest in biosensing and medical diagnostic applications where only relative intensity images are used for the detection/analysis.

*5.2. Microlens-assisted DHM*

As previously stated, another strategy for resolution enhancement in DHM is by using particles in the close proximity of the specimen. Here, we will not deal with randomly-moving particles but with the possibility to locate a microlens-based element in close proximity to the sample. Typically, the microlens is placed on top of the sample and provide resolution enhancement by capturing additional propagation modes that will not be transmitted through the conventional microscope lens without the microlens or, in other words, by transforming the object wave from the higher frequencies to the lower ones.

We have established two microlens modalities inside this SR strategy. As the first modality, microsphere-based microscopy systems [385-387] have garnered lots of recent interest from the research community because of its simplicity and effectiveness for focusing light and imaging beyond the Abbe's diffraction limit [388-395]. The key concept in microsphere-assisted microscopy [387] is really close to the idea of a solid immersion lens (SIL) proposed back in the 90s [396, 397]. In the SIL concept, a hemispherical solid immersion lens increases the refractive index value of the object space from the air ($n_{air}$) to the material used to build the solid immersion lens ($n_{SIL}$). This fact means that the smallest feature that can be resolved has a lateral size of $\lambda/(2\ n_{SIL}\ NA)$, approximately. Microsphere-based assisted microscopy replaces the plano-convex SIL by a full spherical lens and provides higher intrinsic flexibility, better resolution, higher magnification, and longer working distances than SIL microscopy [398].

This new type of microsphere-based microscopy has been mainly successfully applied to intensity-based microscopy [388-395, 399] but also to DHM [400-407]. Figure 25 includes both the experimental layout and some results obtained by using the technique reported in Ref. [401]. Coherent illumination coming from a laser source is spatially decorrelated by a rotating diffuser to remove speckle noise. The illumination beam reaches the sample plane after passing through a Mirau microscope lens (Nikon, 0.3 NA, 10X) and a silica microsphere (refractive index of 1.46 at the laser wavelength). The microsphere is mounted at the end of an optical fiber (acting as a holder) and its position is controlled by a XYZ micro-positioner. The reflected light from the sample surface acts as an object wave, and the reflection from the incorporated glass-plate in the Mirau microscopic objective acts as a reference wave. The interference pattern [Fig. 25(b)] between the two beams is recorded by a digital camera and off-axis configuration is allowed by tilting the sample just a small angle (smaller than 10º). From this hologram it is possible to achieve superresolved QPI of the inspected RBC [Fig. 25(c)]. The authors of the study proposed the method for 3D imaging and identification of biological samples. In particular, they were able to distinguish RBCs coming from a healthy subject and from a subject who tested positive for thalassemia minor.

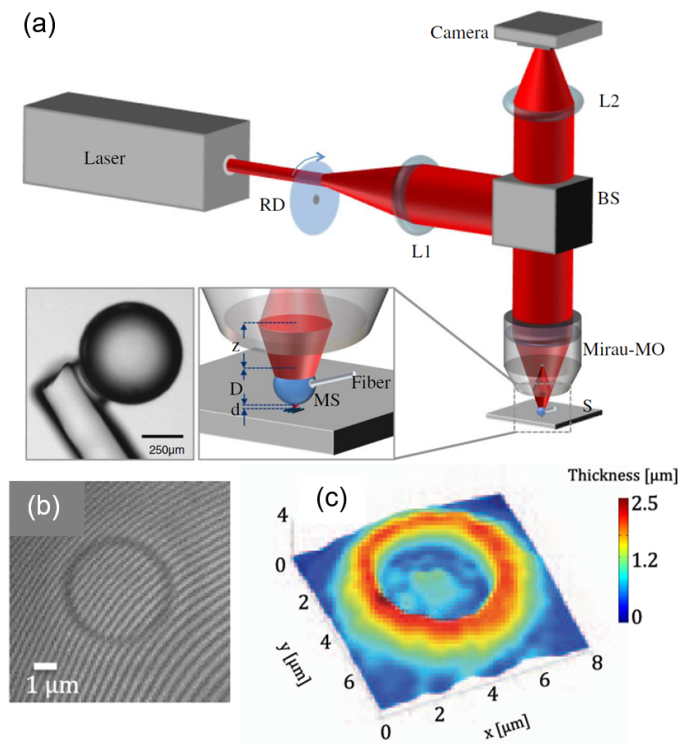

Figure 25. Microsphere-assisted super-resolved Mirau DHM: (a) scheme of the experimental layout, (b) hologram of a RBC considering the proposed technique, and (c) 3D phase map of a RBC. Images taken from Ref. [401].

However, one of the main drawbacks of microsphere-assisted superresolved DHM is the very restricted FOV provided by the method since, essentially, the FOV is limited by the size of microsphere which is used in the experiments. Thus, the useful FOV is approximately reduced to 10x10 times in XY directions in comparison with the same optical layout without the microsphere [401].

The second type of the microlens-assisted SR modality overcomes this reduced FOV imposed by microsphere-assisted DHM by providing a larger FOV even in comparison with conventional DHM. The reason comes from the use of a lensless imaging configuration, as previously introduced in Section 4.1, provides a huge FOV since the whole digital sensor area is essentially the useful FOV. This second modality is usually named as wetting film on-chip microscopy invented a few years ago [408]. After that, several contributions have been reported to be able detect individual sub-100 nm particles and viruses ~100 nm over a FOV of >20 mm$^2$ [409-412]. In these techniques, lateral shifting between illumination and sample provides pixel super-resolution, resulting in the visualization of sub-resolution particles. Wetting film microscopy has also be combined with imaging lenses in a similar way that microsphere-assisted DHM for the detection and sizing of single nanoparticles (100 and 200 nm) over a FOV of 19.1 mm$^2$ [413].

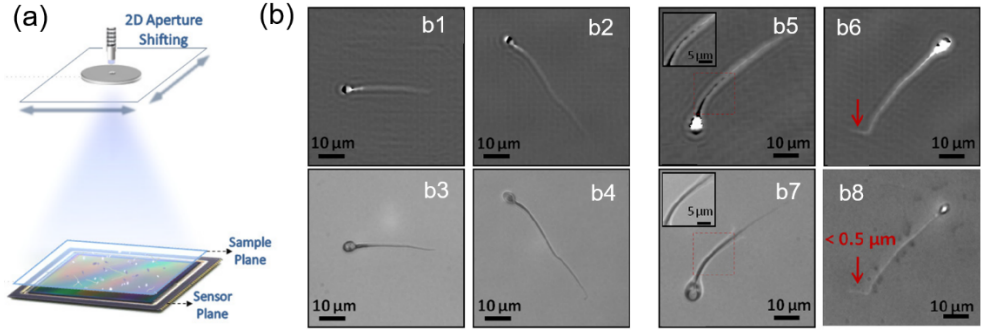

Figure 26. Wetting film superresolved on-chip microscopy: (a) scheme of the experimental layout, (b) experimental results of a sperm cell sample: (b1-b2) phase reconstructions obtained from pixel SR algorithm in lens-free microscopy, (b3-b4) comparative images from a 60X 0.85NA objective-lens, (b6-b6) reconstructed phase from wetting film superresolved on-chip microscopy, and (b7-b8) comparative images from a 60X 0.85NA objective lens. Images taken from Ref. [409].

Figure 26(a) shows schematically the experimental layout of the wetting film microscopy Similar to a classical LHM configuration, a sample is located on top of the digital sensor. The difference lies on that, wetting film on-chip microscopy is provided by forming individual micro-droplets from an initial droplet load which is dissolved, agitated, and spread by mechanical vibration over the sample. In order to increase the resolution of the basic LHM layout, pixel SR algorithm in lens-free microscopy is applied [271, 272, 274]. Thus, the comparison [Fig. 26(b)] is performed in terms of the reconstructed phase from: (b1-b2) the pixel superresolved lensfree microscopy platform, (b5-b6) the wetting film pixel superresolved lensfree microscopy concept, and (b3-b4)-(b7-b8) the comparative images obtained by a 60X 0.85NA microscope objective. Thus, the resolution enhancement of the phase images over the regular superresolved images is directly noticeable by comparing Figs. 26(b1-b2) and (b5-b6). As an example, the end of the sperm tail marked with a red arrow in Fig. 26(b8) measures < 0.5 μm in width and it becomes faithfully imaged using the wetting film superresolved LHM as illustrated in Fig. 26(b6).

## 5.3. Evanescent field based techniques for QPM

Scalar diffraction theory, considering the angular spectrum of plane waves, provides us with a very useful formulation to understand the propagation of the complex field from a given plane (usually represented by $z = 0$) to another plane at z distance [414]. Using this linear framework, the complex field distribution generated by an input complex transmittance (e.g., a sample) illuminated by a monochromatic plane wave can be understood using Fourier analysis across any plane. In particular, an input plane amplitude distribution $U(x, y; 0)$ and the propagated plane $U(x, y; z)$ is linked by the various spatial Fourier components, which can be identified as plane waves traveling in different orientations from the input plane to the output plane. And the complex amplitude of the field at z distance can be calculated by adding coherently the contribution of these plane waves at z distance, taking into account the phase-shifts caused by propagation. This can be mathematically expressed as:

$$U(u,v,z) = \iint_{-\infty}^{+\infty} A(u,v,z) \exp\left[j2\pi(ux+vy)\right] du\, dv \tag{9}$$

where $A(u, v, z)$ is the angular spectrum of plane waves at z distance which, in turn, comes from the angular spectrum of plane waves $A(u, v, 0)$ at z = 0 after propagating a distance equal to z. $A(u, v, z)$ and $A(u, v, 0)$ are related by the Helmholtz equation, and can be expressed with:

$$A(u,v,z) = A(u,v,0)\exp\left(j\frac{2\pi}{\lambda}\sqrt{1-(\lambda u)^2 - (\lambda v)^2}\, z\right) \tag{10}$$

where $\lambda$ is the wavelength of the monochromatic plane wave.

Equation (10) establishes the relation between the angular spectrum of plane waves arriving at a plane at z distance from another plane taken as origin (z = 0) and it can be divided into two different regimes: i) $(\lambda u)^2 + (\lambda v)^2 < 1$ and ii) $(\lambda u)^2 + (\lambda v)^2 > 1$, both mathematically and physically possible. The first case yields in a complex argument for the exponential function, so it represents real propagating waves traveling from one plane to another. We will refer to these components as propagating waves. In this case, the effect of propagation over a distance equal to z is simply a change of the relative phases of the different components of the angular spectrum (those ones satisfying the first condition). The different phases come from the fact that each plane wave propagates at different angle, so each will travel a different distance originating a relative phase delay when considering other components.

However, the second case produces a real argument in the exponential function and Eq. (10) can be rewritten as

$$A(u,v,z) = A(u,v,0)\exp(-\sigma z), \tag{11}$$

where $\sigma = \frac{2\pi}{\lambda}\sqrt{1-(\lambda u)^2 - (\lambda v)^2}$. These wave components are rapidly attenuated in propagation and will not exist in the far field since they become exponentially attenuated in the near field (a distance comparable to the wavelengths in that medium). They are usually called as evanescent waves.

The previous two regimes establish the first truncation in the wave components that, being present in $A(u, v, 0)$, will not be considered in Eq. (10) for the definition of $A(u, v, z)$. It is usually assumed that propagation introduces a linear spatial filter having a simple transfer function $H(u, v)$ in the form of

$$H(u,v) = \begin{cases} \exp\left(j2\pi\frac{z}{\lambda}\sqrt{1-(\lambda u)^2 - (\lambda v)^2}\right), & \sqrt{u^2+v^2} < \frac{1}{\lambda} \\ 0, & otherwise \end{cases} \tag{12}$$

So, thinking in terms of resolution, the more components we will be able to transmit from z = 0 to z distance the better the resolution limit will be. However, there are some limitations in the maximum achievable resolution incoming from the spatial filtering provided by the propagation transfer function: firstly, evanescent waves will not take part in the complex field at z distance under free space propagation. And secondly, the number of propagating waves arriving at z distance will be lower than the ones generated by diffraction at $z = 0$ because of the presence of additional diaphragms that the complex field will find during propagation. In other words and with imaging systems in mind, the number of propagating components will be again limited by the aperture size of the imaging lens used to form the final image. This is a second limitation in the number of components capable of travel from an object to its image, so this is a second limitation in image resolution. Sections 3 and 4 deal with approaches aimed to enhance the resolution limit by applying specific methods for retrieving a higher number of diffracted waves than the same imaging system without those approaches. This type of methods allow superresolved imaging but restricted to propagating waves.

Now, the obvious question to be answered is: is it possible to further improve the resolution limit by considering evanescent waves? The answer to this question is positive and, although there are more methods such as scanning near-field optical microscopy and the use of different types of "superlenses", the aim of this paper is to review SR approaches for QPI so we will include in this section some approaches capable of QPI by using evanescent waves.

### 5.3.1. Solid-immersion imaging interferometric nanoscopy

Immersion techniques are widely known for maximizing the resolution in optical microscopy. Assuming that the resolution limit for an air-immersed imaging system is proportional to $\sim\lambda/NA_{air}$, immersion methods increases the resolution by a factor equal to the refractive index of the immersion medium ($n_{im}$) in the form of $\sim\lambda/NA_{im}$ where $NA_{im} = n_{im} NA_{air}$. But the use of immersion objectives are sometimes awkward due to practical issues and excluded depending on the application. For this reason, oblique beam illumination can provide similar synthetic NA values to immersion objectives but working with more modest air-immersed lenses [119].

Oblique beam illumination has been pushed to its limits by the group of Prof. Brueck in Alburquerque [103, 182, 415]. The technique, named as imaging interferometric nanoscopy, is an attractive alternative that takes advantage of SA generation, using multiple illumination and collection directions, to synthetically expand up the spatial-frequency domain coverage provided by the objective lens. Using this strategy, the authors demonstrated a resolution limit close to the minimum theoretical one of $\sim\lambda/2$ for air-immersed imaging systems [103] and to $\sim\lambda/2n_{im}$ for immersion media [415]. In other words, their work showed resolution limits reaching the linear spatial filter imposed by propagation (Eq. 11).

However: is it possible to go further (or lower) in terms of resolution with oblique beam illumination? Again, Brueck et al. demonstrated SR imaging by SA microscopy with evanescent wave illumination [170]. As a main difference with previous air-immersed approaches, the sample is placed on top of a transparent high-index substrate and evanescent wave illumination of the sample is allowed by extreme off-axis illumination with illumination propagation beams beyond the total-internal reflection (TIR) angle of the transparent substrate (we will go in depth with this approach in next section). Thus, the sample scatters the evanescent illumination, coupled through the substrate, into propagating diffracted waves that are collected in a standard, full-field DHM configuration. In this configuration, the technique only uses the immersion medium for providing the evanescent illumination while the space between the sample and the imaging objective remains air (this is why the authors named it as effectively half-solid immersion imaging interferometric nanoscopy). But it can be defined as a full immersion collection scheme by placing the imaging lens on the same side than the substrate [415]. Both modalities generate the technique named as solid-immersion imaging interferometric nanoscopy (SIIN).

Figure 27 includes some experimental results validating the proposed SIIN method. The experimental setup scheme [Fig. 27(a)] shows only the sample illumination stage (the required reference beam for the interferometric recording is not shown) for 3 different cases: A – regular imaging conditions with the objective normal to the substrate surface (resolution limit $\sim\lambda/NA$), B - objective tilted away from the optical axis to access a higher diffracted content by the sample (resolution limit up to a maximum of $\sim\lambda/2$ for maximum air-immersed NA and oblique beam illumination), and C – objective on the side of the substrate with grating [SIIM approach with a resolution limit $\sim\lambda/2(n_{im}+1)$]. The evanescent waves from the higher frequency content of the object are coupled back into the substrate by the boundary conditions at the substrate interface and propagate in the substrate at angles beyond the angle for total internal reflection (TIR). For a flat interface, the information at these spatial frequencies is not accessible, but the scattered light can be decoupled by a grating on the side of the substrate opposite to the object and redirected to an objective on the grating side. The interferometric recording allows retrieving complex amplitude distribution of the different recordings containing different frequency apertures from which a SA can be synthesized. The composed object images are then obtained by inverse FT of the SA showing SR effect. Figure 27(b) shows the Manhattan test pattern used to show SIIN where the pattern is scaled to different dimensions for the present experiments. In particular, Figs. 27(c)-(d) show the reconstructed images for a line width of 180 and 150 nm, respectively. Considering He-Ne illumination wavelength (633 nm), glass substrate ($n_{im}$ = 1.5) and a 0.4 NA objective lens, the resolution limits corresponding with A, B and C imaging conditions are ~1.58 μm, ~316 nm and ~126 nm, respectively. So, experimental validation of SIIN with evanescent wave illumination was accomplished. Note that, although SIIN was not demonstrated for QPI (only amplitude test images were reported), the interferometric nature of the method enables its use in QPM.

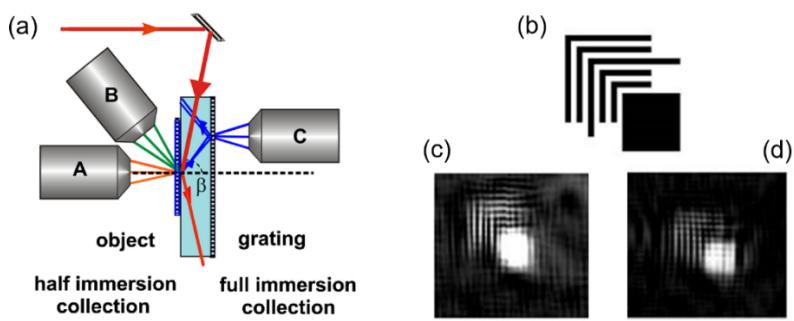

Figure 27. (a) Illumination and collection configurations: A – objective normal to the substrate surface, B – objective with tilt away from the optic axis, C – objective on the side of the substrate with grating; (b) Manhattan test pattern used in the experiments, and (c)-(d) reconstructed images of 180- and 150-nm line widths patterns, respectively. Images adapted from Refs. [170, 415].

### 5.3.2. Total internal reflection holographic microscopy

As previously commented, one way to produce evanescent wave illumination at the sample to be inspected is by using Total internal reflection (TIR). When light passes from inside a higher refractive index $n_1$ medium into a lower-index $n_2$ one and the angle of incidence is greater than the critical angle given by $\theta_c = \sin^{-1}(n_2/n_1)$, all of the light is confined into the first medium by TIR. This well-known effect means there is no propagating field in the second medium except for the evanescent waves. As we have previously stated, the evanescent field is non-propagating and its amplitude decays exponentially over a distance of tens or hundreds of nanometers. But it can be used to modulate the phase of the reflected $n_1$ wave either through inhomogeneous refractive indices in the $n_2$ medium [416], frustrated TIR geometries [417], or to fluorescent TIR microscopy by exciting suitable dye molecules placed in the evanescent field [418].

In 2008 Ash and Kim combined for the first time a TIR system with DHM yielding in the definition of total internal reflection holographic microscopy (TIRHM) [419]. TIRHM appears as an improvement over interference reflection microscopy (IRM) by alleviating some of its drawbacks. In IRM, light waves reflected from two surfaces of the cell substrate interface produce interference fringes which can be analyzed for estimation of the interface thickness profile [420, 421]. But the interference image of the interface is usually complicated by the reflection image of the cell body and its contents. TIRHM utilize a clean reference beam for the holographic recording under off-axis [419, 422] as well as on-axis [423] configurations, one with pros and cons over the other. Since then, TIRDHM has been proposed and validated for the measurement of refractive index distributions [416, 424], simultaneous measurement of refractive index distribution and topography of samples [425, 426], as well as for biological studies [427].

However, TIRDHM is a synthesis of evanescent wave illumination with DHM being developed for real-time QPM of the interface of biological cells moving over a glass substrate. It is useful for obtaining QPI of micron scale surface profiles in near-field with an optical axial resolution of a few nanometers. But the lateral resolution is restricted to the used imaging lens since the evanescent wave components are not transferred to the far field as in the previous section. For this reason, it is unrealistic to expect lateral resolution enhancement using TIRDHM unless other techniques are combined with it.

In recent years, novel methods have been proposed to break the diffraction limit and achieve optical SR imaging in TIR microscopy. Hao et al. proposed a passive spatial frequency shift mechanism for achieving SR imaging in TIR [428]. Essentially, when the sample containing sub-wavelength details is illuminated by the evanescent waves, these details can strongly interact with the field (scattering and diffraction), and a passive frequency shift happens. This fact means that the image generated in the far field will contain the information that originates from the offset region in Fourier space, offset region generated by the evanescent wave illumination. So similarly to oblique beam illumination, SR is obtained only along the direction that is parallel to that of the illumination light. Hence, it is necessary to collect a sequence of images with different illumination orientations to achieve SR in both dimensions. Figure 28 includes the experimental results obtained by the proposed evanescent wave induced frequency shift approach. The test object [Fig. 28(a)] contains feature sizes of 130 nm linewidth with 130 nm gap width in three different directions. Those details are smaller than the diffraction limit provided by the microscope objective (Nikon 100X 0.8NA lens) under white light illumination [Fig. 28(b)]. However, the proposed technique retrieves a single frame image [Fig. 28(c)] for any fixed illumination direction that resolves the input object [Fig. 28(d)] after using simple digital post-processing involving FT of the recorded image, Fourier filtering, backward frequency shift to restore the initial frequency distribution, and inverse FT. In order to achieve 2D SR, the illumination excitation should be rotated to different directions [Fig. 28(e)].

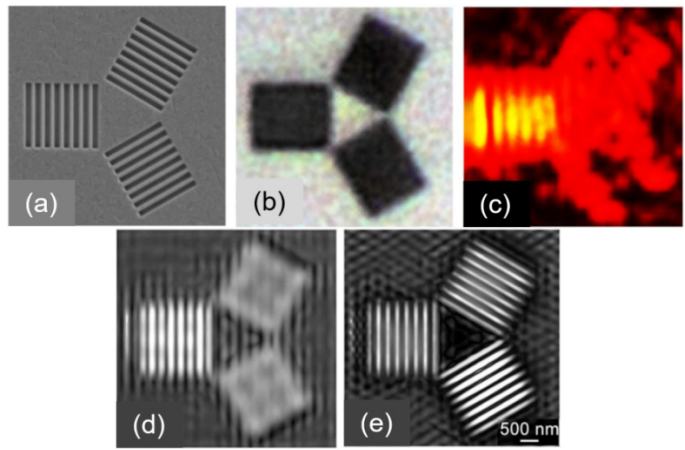

Figure 28. SR imaging of TIRHM by evanescent-wave-induced frequency shift. As a comparison, the images of the sample are taken by (a) focused ion beam microscopy, (b) a general white-light optical microscope, (c) direct image provided by the proposed method with a fixed illumination direction, (d) direct image after the needed image processing (backward frequency shift and inverse FT), and (e) final reconstructed image considering 3 different illumination directions. Images taken from Ref. [428].

A similar procedure has been applied to TIR dark-field microscopy [429]. Many coherent images caused by evanescent wave illumination coming from different directions are incoherently superposed to resolve objects that remain unresolved by conventional TIRF (fluorescence) microscopy. Experimental validation was reported for 190 nm beads but the technique, although not useful for QPI, seems to be promising label-free SR imaging in live-cell microscopy.

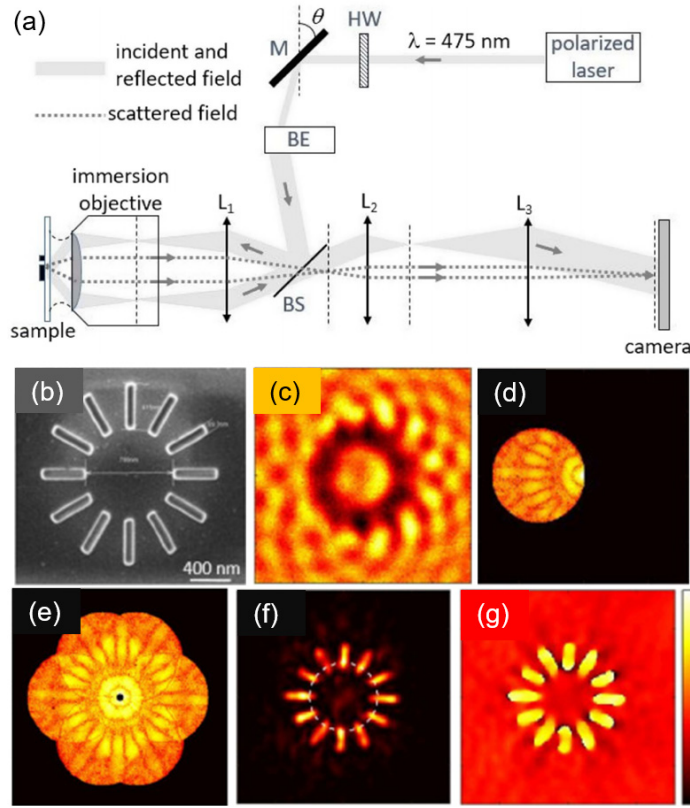

Figure 29. SR phase imaging via TIR DHM: (a) scheme of the experimental layout, (b) test sample, (c) conventional intensity image obtained at normal incidence in non-TIR mode, (d) one of the 6 retrieved elementary apertures in oblique beam evanescent wave illumination TR DHM, (e) the generated SA, and (f)-(g) the superresolved images in dark field intensity and quantitative phase (scale bar is phase in rads) visualization modes, respectively. Images taken from Ref. [430].

TIR microscopy has also be combined with several illumination angles in order to achieve SA generation to finally get reconstructed imaged with improved resolutions compared to standard label-free microscopy techniques [430]. The core idea of the technique is that the intensity detected in TIR microscopy can directly give access to both the phase and the amplitude of the field scattered by the probed sample. This is achieved by assuming that the beam reflected by the TIR interface is very strong in comparison with the back reflected light scattered by the sample. This assumption is very similar to the Gabor idea of holography where the sample provides a small perturbation of the transmitted beam. So, the back-reflected beam by the TIR interface acts as an off-axis reference wave for conventional DHM. The layout scheme and experimental results are shown in Fig. 29. A supercontinuum light source is filtered by an interference filter centered at 475 nm with a spectral width of 6 nm. A motorized half-wave plate is used to adjust the polarization direction of the linearly polarized illumination wave using to be parallel to the TIR interface for each illumination angle. A fast-steering mirror permits to control the deflection of the beam and a beam expander generates a wide collimated beam. Because of the geometrical design, rotating the mirror varies the illumination angle without shifting laterally the beam on the object. TIR illumination of the sample and collection of the reflected intensity were performed through the microscope objective (Nikon CFI Apo TIRF, 100X, 1.49NA). The signal is imaged on a digital camera with a global magnification of about 290X by using additional relay lenses.

The setup was used on calibrated resin star samples deposited on a glass slide [Fig. 29(b)]. The inner diameter of the star is 800 nm and each of its 12 branches has a width of 90 nm and a length of 400 nm. The sample has a height of 160 nm. Figure 29(c) shows the intensity image obtained at normal incidence in non-TIR mode where the sample is clearly unresolved. The resin star was illuminated in TIR along six directions of polar angle about 60°, and with azimuthal angles regularly spaced every 60°. Figure 29(d) shows one of the six retrieved shifted apertures from which the SA [Fig. 29(e)] is generated after some digital processing. Finally, the SR image of the star sample can be reconstructed in intensity (dark field imaging mode, Fig. 29(f) and phase [Fig. 29(g)] by 2D inverse FT of the generated SA. The reported approach [430] defines a common-path TIRDHM working with SR capability incoming from oblique beam evanescent wave illuminations for SA generation.

### 5.3.3. Surface plasmon holographic microscopy

The first observation of surface plasmons is reported by Wood [431] when he used a metallic grating to study the diffraction of polychromatic light and found an unexpected narrow dark band in the spectrum. The first theoretical investigation of surface electromagnetic waves was carried out by Zenneck [432] a few years later. However, experimental validation on surface plasmon resonance started when Otto [433] and Kretschmann and Raether [434] demonstrated optical excitations of surface plasmons using different configurations of attenuated total reflection method. These pioneering works had formed the foundation of optical biosensors based on surface plasmon resonance and stimulated numerous researches in subsequent decades [435]. Surface plasmons are electromagnetic surface waves that propagate along the interface between conductors and dielectrics. Their significance resides in the extremely high sensitivity of surface plasmons to small changes in the dielectric properties of the substances that are in contact with the conductors. This characteristic enables surface plasmons for biosensing. Surface plasmon microscopy (SPM) [436] uses the evanescent field in surface plasmon resonance as the illumination light. Initially, surface plasmon resonance sensors were based on light intensity detection [437] but phase sensing methods have been rapidly attracted the attention in recent years because of the higher detection sensitivity [438-443]. In particular, SPM has been combined with DHM generating surface plasmon holographic microscopy (SPHM), a technique capable of simultaneously obtain amplitude- and phase-contrast surface plasmon resonance images [444-446]. SPHM has been experimentally validated in quantitatively monitoring of tiny refractive index variations [447], biological tissue imaging [448] and mapping thin film thickness in near field with high sensitivity and temporal stability [449], just to cite some examples.

Figure 30 shows SPHM proposed by Ref. [449]. A linearly polarized laser beam with its polarization direction having $45°$ with respect to the horizontal direction is used to illuminate the sample allocated on a coverslip coated with 50 nm gold film and the SPR occurs with a specific incident angle $\theta$ that can be controlled by changing the focal position $d$ on the back focal plane of the objective. Being illuminated by the evanescent field and the back-reflected beam is imaged by a high NA oil immersion microscope objective (Nikon Plan Apo Lambda DM 100X NA 1.45) to the CCD sensor. To allow holographic recording, a common-path interferometer based on a Wollaston prism is assembled before the digital camera. Essentially, the incoming beam is split into $s$- and $p$-polarized components by the prism. At the resonance angle, the reflectivity of the $p$ wave drops drastically (mainly reflected by the gold film) and the phase exhibits a marked jump. Conversely, the $s$ wave does not excite the SPW, and therefore its reflection spectrum does not demonstrate an absorption dip or a significant phase variation [450]. Both beams interfere with each other after passing through the polarizer *P* for fixing a common polarization state. By analyzing the phase difference between the two waves, the authors measured the thickness of a multilayer grapheme film (material with absorption). As in shown in Figs. 30(b)-(c), the retrieved thickness profile perfectly agrees with the one retrieved by AFM. As a consequence, the film thicknesses of these two kinds of materials in near field have been determined unambiguously by using SPHM without a priori information. Theoretically, the measurement resolution can reach sub-nanometer level and the measurement range is around as 0 ~ 50 nm.

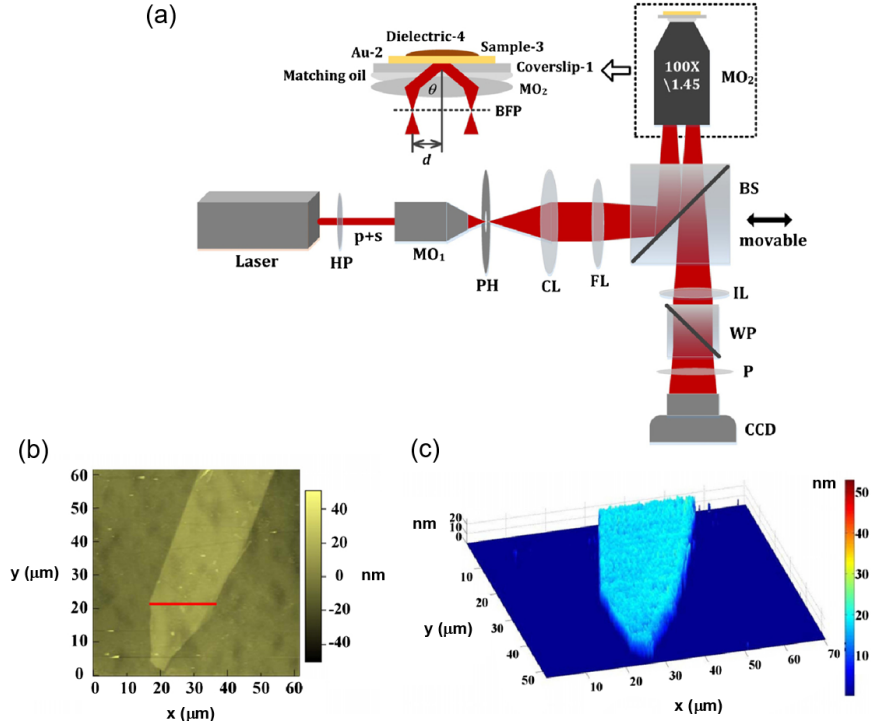

Figure 30. Surface plasmon holographic microscopy. (a) Experimental setup for SPHM, (b)-(c) AFM image and reconstructed 3D thickness distribution by SPHM, respectively, of the multilayer graphene film. Images taken from Ref. [449].

Similarly to TIRDHM, SPHM presents a high surface sensitivity compared to conventional optical microscopy and it is very useful for exciting sub-wavelength details of the sample attached to the interface. This allows resolution enhancement in the axial direction and offers the unique ability to monitor in real-time and without any labeling micrometer-size biochemical phenomenon on wide field-of-views. This fact is very useful for measuring tiny 3D thickness and/or refractive index distributions. But leaving aside geometrical aberrations introduced by the prism, the lateral resolution is essentially defined by the imaging lens used in the experimental setup [451, 452]. Although some approaches for achieving diffraction limit imaging and go beyond (SR imaging) have been reported [453-455], those methods are related to SPM and not with SPHM.

Recently, a novel microscopy method combining surface wave illumination and FPM algorithm has been proposed to achieve SR in QPI [456]. By surface wave illumination the authors referred to evanescent and surface plasmon waves that are simultaneously excited through an oil-immersion objective lens. Thus, the projected far-field images are scattered waves that carry sub-wavelength information of the specimen incoming from the interaction between the surface waves and the sample. By digital post-processing of the recorded images using a FPM algorithm [53], the intensity and phase distributions of the sample are recovered yielding in an image with enhanced resolution and improved contrast. The underlying principle can be understood as illumination of the sample with surface waves propagating along different directions, which shifts the sample's high-frequency bands into the pass-band of the imaging system. Then, an expanded spectrum of the sample is recovered by synthesizing these frequency bands.

Figure 31 shows the experimental setup for the surface wave illumination Fourier ptychographic microscopy (SWI-FPM) technique. A laser beam ($\lambda$ = 640 nm) is used to sequentially illuminate the sample through an oil-immersion objective (1.49-NA, 100X) under different illumination directions by using a 3D galvanometric mirror unit. The mirror unit is aimed to, aside of varying the 2D incidence direction, change the radius of the scanning circle in order to adapt the illuminating angle depending on which type of surface wave is being stimulated. Thus, TIR evanescent waves are excited for illuminating angles larger than the critical angle of the evanescent wave. Therefore, surface plasmon waves are stimulated when the illuminating angle equals to the excitation angle for plasmon waves (condition given by $k_0 n_{ill} \sin\theta = k_0 (\varepsilon_{Ag}\varepsilon_{air})/(\varepsilon_{Ag} + \varepsilon_{air})$, being $\varepsilon_{Ag}$ and $\varepsilon_{air}$ the permittivity of silver and air). Finally, the generated surface waves interact with the sample leading to a passive frequency shift induced in the Fourier domain [428]. The scattered waves projected to the far field are collected back by the same objective lens, and imaged onto two digital sensors for the recording of the sample's information in the spatial and the Fourier domains.

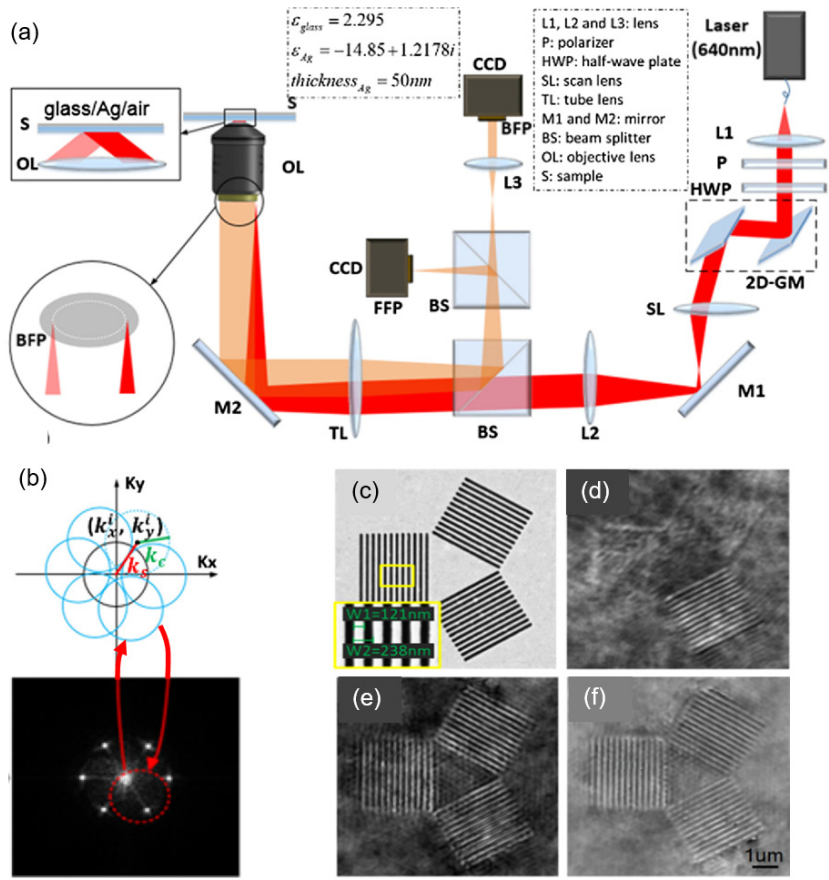

Figure 31. Surface wave illumination Fourier ptychographic microscopy (SWI-FPM). (a) Experimental setup; (b) theoretical (up) and experimental (down) generated SA; (c) test sample used in the experiments (120 nm line width); (d) intensity image taken using an illumination angle of 72°; (e)-(f) the intensity and phase retrieved images, respectively, after applying SWI-

FPM. L1-L3, lens; P, polarizer; HWP, half-wave plate; SL, scan lens; TL, tube lens; M1 and M2, mirrors; BS, beam-splitter; OL, object lens; S, sample. Images adapted from Ref. [456].

A grating sample consisting in 3 gratings of 120 nm line width is used in the experimental validation. The authors captured seven images, one under normal on-axis illumination and the other six images from surface wave illumination at six different propagating directions that are controlled by the 2D Galvanometric mirror unit. Fig. 31(b) depicts the theoretically (up) and the experimentally (down) generated SA as a consequence of the addition of all the recovered band-pass apertures. Figure 31(c) depicts the test sample while Fig. 31(d) includes the retrieved intensity image when using surface wave stimulation with an illumination angle of 72°. There are two grating elements that are not resolved since their spectral information is contained in a different direction but there is one element perfectly resolved by the proposed SWI FPM method. After the whole set of images are recorded and the FPM algorithm applied, a SR image is obtained not only in terms of intensity [Fig. 31(e)] but also the phase distribution [Fig. 31(f)]. The grating structure at azimuth $90^o$ and $135^o$ cannot be resolved in Fig. 31(d), but they are well resolved in Fig. 31(e)-(f), implying a significant resolution enhancement has been achieved with the proposed method.

In addition to resolution enhancement, Ref. [456] also presents a comparison about the image contrast provided by TIR evanescent wave and surface plasmon wave illuminations (Fig. 32). The images in Figs. 32(a1) and 32(b1) are taken under evanescent wave illumination coupled by s-polarized light with an incident angle of 43.1°. The images in Figs. 32(a2) and 32(b2) are captured under surface plasmon wave illumination coupled by p-polarized light with the same incident angle. Figures 32(c1) and 32(c2) are the recovered intensity images using the proposed method. As the main outcome, a field enhancement effect of the surface plasmon wave can be clearly observed both in the raw data and the recovered data. This fact is made evident by plotting [Fig. 32(d)] a profile of the recovered intensity curves for both surface wave illumination modalities and yielding in an average contrast of 0.64 and 0.73 for TIR evanescent waves (green plot) and surface plasmon waves (red plot), respectively. The plot demonstrates a higher contrast improvement in the retrieved intensity images by surface plasmon waves than for TIR evanescent waves.

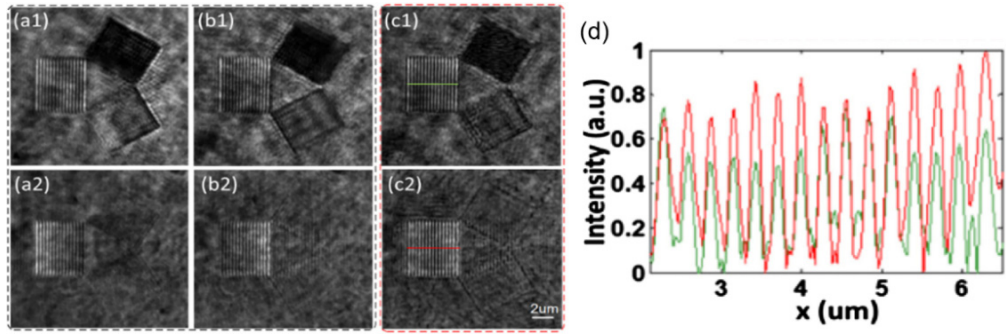

Figure 32. Comparison of the image contrast on a 150 nm linewidth gratings achieved by the evanescent waves and the surface plasmon wave illuminations: (a1) and (b1) Raw images taken by counter-propagating evanescent wave illumination. (c1) Intensity image reconstructed with (a1) and (b1). (a2)–(c2) Corresponding images under surface plasmon wave illumination. (d) Recovered intensity profiles alone the color-coded lines in (c1) and (c2). Images taken from Ref. [456].

It is worth insisting on the fact that the methods based on evanescent waves, total internal reflection or surface plasmons rely on the extreme proximity of the inspected volume of the sample with an interface between media. This way the axial resolution is extremely high, at the price of being limited to the close vicinity to the interface. As a consequence, the methods cannot be in general applied to arbitrary imaging conditions. In the case of surface plasmons, the condition of a dielectric-metal interface is a serious drawback for many practical implementations.

## 6. CONCLUSIONS

In this paper, we have reviewed the state-of-art in resolution enhancement approaches in coherent imaging, with the main focus on phase imaging of microscopic objects. For lensless QPM, SR is attained by synthesizing a larger hologram coming from an expanded aperture or with a smaller effective pixel size. This strategy can be implemented by relatively shifting the sample and the CCD camera, tilting the illumination or inserting a grating in the setup. For lens-based QPM, SR is achieved by synthesizing a larger synthetic/effective NA than the one defined by the conventional objective lens. For this purpose, oblique illumination, SI and speckle illumination were proposed to improve the spatial resolution of the final image.

Aside of QPI via interferometric approaches, phase imaging can also be implemented in absence of reference wave (techniques termed as reference-less phase imaging). The key of this type of imaging techniques is either generating a reference wave from the object wave itself or coding the phase variation into intensity variations by applying different constraints, such as recording the intensity pattern of an object wave in the different axial planes, different wavelengths or different aperture tailoring. The resolution of these reference-less QPM approaches is limited by the same factor as in interferometric approaches. Therefore, all the illumination schemes used for SA in interferometric based phase imaging approaches can be extended to reference-less phase imaging approaches.

All those methods are included in Sections 3 and 4 and, essentially, are based in multiplexing the spatial-frequency content of the object spectrum into a different degree of freedom which is not used in the system. The underlying principle of this generic multiplexing is addressed in Section 2 yielding in a SPB adaptation process between the input object and the imaging system. We have also extended our review to other approaches where SR effect is achieved in a different way (Section 5). SGH DHM exploits the non-linear behavior of some materials for generating second harmonic waves, thus reducing the illumination wavelength at a half and improving the resolution limit by a factor of 2. Other methods are capable of capturing additional propagation modes that only propagates in the close proximity of the sample. These modes are converted from the evanescent to the propagating field by using microlenses in the close proximity of the sample or by using the physics behind total internal reflection (TIR DHM) or surface plasmon resonance (SPHM).

Under a different perspective, Sections 3 and 4 considers that the key of resolution enhancement approaches is to synthesize a larger aperture yielding in an effective NA value of $NA_{eff} = NA_{illum} + NA_{system}$. As a result, the highest resolution limit we can achieve is restricted to the physical limit $\lambda/2$ imposed by the Abbe's diffraction limit. However, Section 5 shows that SR above the Abbe limit for coherent imaging is also possible. There are additional methods aimed to this. For instance, superoscillation-based imaging has been proved to surpass the diffraction limit, considering that a wavefront oscillates faster than its highest constituent frequency component over a finite interval [457, 458]. Furthermore, the use of random movement of nanoparticles in the close proximity of the sample has also been proposed as a way to achieve SR by scanning and localizing their positions with sub-resolution accuracy [459-461]. However, those methods are not (at this time) connected with QPM and we have decided not to include them in this review.

SR for coherent imaging is of great significance to understand the nature of light in terms of its phase, especially when the light propagates through scattering mediums. Therefore, SR approaches for coherent imaging are meaningful since they extend significantly the intrinsically-limited SBP of the system. The approaches we reviewed here provide a way to improve the resolution and keep the same FOV, therefore improving SBP significantly. To allow this, another degree of freedom needs to be sacrificed (mostly time). Nevertheless, with the fast development of field modulation techniques (intensity and phase modulation) and imaging devices (such as high-speed CMOS/CCD cameras), the time needed for high SBP through-out imaging will be shortened significantly. In the meantime, multiplexing techniques also provide a way to realize single-shot phase retrieval with SR [367].

**FUNDING**

This research was partially supported by National Science Foundation of China (NSFC) (61475187, 61605150, U1304617). This work was also supported by the Spanish Ministerio de Economía y Competitividad and Fondo Europeo de Desarrollo Regional under the projects FIS2007-60626, FIS2010-16646, FIS2013-47548-P and FIS2017-89748-P. Part of the project was funded by China Postdoctoral Science Foundation (2017M610623), and the Fundamental Research Funds for the Central Universities under Grants No. JB160511, XJS16005 and JBG160502.

**ACKNOWLEDGEMENTS**

*Short author biographies (in order of coauthorship)*

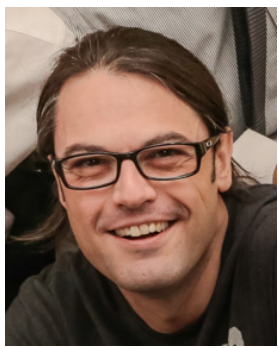

**Vicente Micó** received the MS degree in Physics from the Universitat de Valencia in 1999 and the BS degree in Optics and Optometry in 2000. He obtained his Ph.D. in Physics in 2008 at the Universitat de Valencia. He is currently Associate Professor at the University of Valencia

He is member of the Optoelectronic Image Processing Group headed by Prof. García. According to his university degrees, his research interests are in two areas: opto-electronic imaging and optometry/vision science. As a physicist, he is actively working in optical metrology, digital holography, digital holographic microscopy and optical superresolution, including vibration and deformation metrology, speckle applications, lensless coherent imaging and setups based on spatial light modulators. As an optometrist, he is working in theoretical aspects on physiological and visual optics as well as in the development of novel optometric instruments for the measurement of ocular parameters. He is coauthored of around 100 peer-review papers (H = 20) as well as around 15 book chapters regarding resolution enhancement in coherent imaging. He is of the Spanish Optical Society and SPIE.

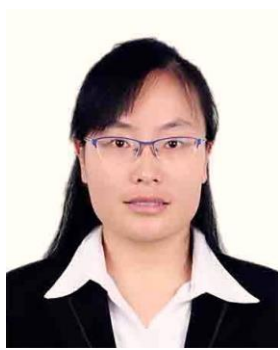

**JuanJuan Zheng** is a lecturer in School of Physics and Optoelectronic Engineering, Xidian University. She received her B.Sc. degree in Physics at Shanxi Normal University, and obtained her Ph.D. in Optics at Xi'an institute of optics and precision mechanics, Chinese Academy of Science (CAS). She conducted the research of phase microscopy and coherent anti-stokes raman scattering (CARS) microscopy at University Jena during 2012-2014. She was awarded "The Excellent Doctoral Dissertation" from CAS (Top 3%) and "The Best Doctoral Dissertation of Optical Engineering Society" in 2016. Her research interest includes (a) Digital holographic microscopy with speckle illumination; (b) Coherent anti-Stokes Raman spectroscopy. She has authored 30 peer-reviewed papers, of which some were selected as "Spotlight on Optics", "Highlights of IOP 2015", "IOP-select", "Top-10 download in Applied Optics".

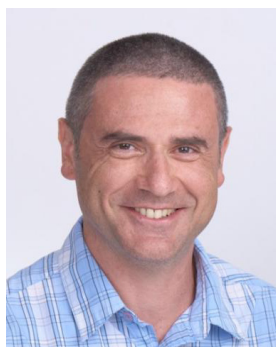

**Zeev Zalevsky** received his B.Sc. and direct Ph.D. degrees in electrical engineering from Tel-Aviv University in 1993 and 1996 respectively. He is a full professor of optics from 2010 and currently, he is a Vice Dean of Engineering at the Bar-Ilan University in Israel.

He is head of the Electro-Optics study program at the Faculty of Engineering and director of the Nano Photonics Center at the Institute of Nanotechnology and Advanced Materials (BINA). The main research fields in his lab are bio-medical optics and bio-imaging, nano-photonics, silicon photonics, super resolution, diffractive optical elements and beam shaping, in-fiber devices, fiber optics, and RF-photonics. He has published 450 peer-review papers (H = 45). He also acted as Editor of 4 books and around 20 book chapters regarding superresolution and quantitative phase imaging. Among others, he received the Krill prize from the Wolf foundation (Wolf prize for young scientists) in 2007, the ICO prize and Abbe's medal for significant contribution in the field of optical super resolution in 2008, the Juludan prize for advancing technology in medicine in 2009, the Winner of The International Wearable Technologies (WT) Innovation World Cup 2012 Prize for the invention of the "Opto-Phone" in 2012, the Tesla Award for Outstanding Technical Communication on Electro-Optics in 2013, the SPIE Startup Challenge Winner Prize for a project of tactile cornea stimulation allowing functionality to blind people in 2017, and The Winner of the Photonics Award (1st place) at Startup World in Munich for a solution for hand held remote and contactless bio-sensing technology (ContinUse Biometrics Ltd.). He is also SPIE, EOS and OSA member and selected as IEEE Senior Member.

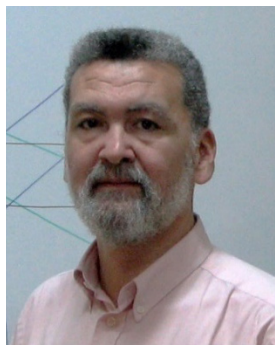

**Javier García** received his B.Sc. and Ph.D. in Physics by the Universitat de Valencia in 1989 and 1994 respectively. Since 2008, he is full Professor in Optics at the Universitat de Valencia, Spain.

He leads the Optoelectronic Image Processing Group at the Department of Optics. His has worked on optical imaging and signal processing, optical pattern recognition, the use of space-frequency based representations by means of the fractional Fourier transform, optical multichannel processing; the investigation of spatial pattern formation in photorefractive resonators. But nowadays, his main research lines are optical superresolution in coherent imaging and biometric sensing by speckle imaging. This field involves the knowhow on Fourier Optics, microscopy, computer generated holograms, digital holography and image processing. He has leaded private and public funded projects, ranging from local to European projects. He has been advisor for several companies, in particular for Lenslet Labs (in optical computing architectures) and PrimeSense (in 3D real time sensing for Microsoft Kinect). Nowadays he is engaged in three startup companies: ContinUse, Cytolovu, and ZSquare centered in biometrics, digital holographic microscopy, and endoscopy, respectively. He coauthors over 150 peer-reviewed papers (H = 24). He has also several patents, three of which are under commercial use. He has been awarded the IS&T's 2015 Image Engineering Innovation Award, the International Wearable Technologies Innovation World Cup 2012, Finalist in the 2015 Berthold Leibinger Innovationspreis, he is member of the Spanish Optical Society, EOS, SPIE and OSA and he is Fellow of the OSA.

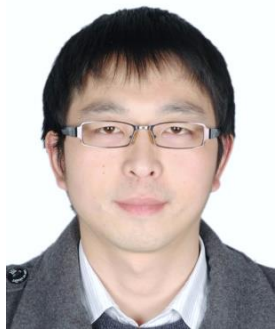

**Peng Gao** studied Physics and received his Ph.D. at the Xi'an Institute of Optics and Precision Mechanics (XIOPM), Chinese Academy of Sciences, in 2011. He was a "Humboldt Fellow" in University Stuttgart (2012-2014) and later Marie-Curie Fellow (IEF) in KIT and University Manchester. He is currently appointed as Principal Investigator (PI) and full professor in Xidian University. He is leading a Biological Imaging and Spectroscopy group, funded by the "Thousand Talents Recruitment Plan" from the government, and Xidian University. His group centers on developing new optical imaging techniques and applying them for biology and industry. The main research lines include but not limited to: (a) Digital holographic microscopy (DHM) and reference-less phase retrieval with modulated illumination; (b) Super-resolution fluorescence microscopy and correlation spectroscopy. He had authored around 50 peer-review papers, including *Nature Photonics*, *Optics Letters*, and *Scientific Reports*. Some of the publications were highlighted by *Nature Methods*, the front cover of *Applied Optics*, and tens of international media such as *Science Daily*, *Physics News*, *Advance of Engineering*.